\documentclass[11pt]{article}
\usepackage{amsmath,amssymb,multirow,jheppub,graphicx, subfigure}
\usepackage{amsthm}
\usepackage{centernot}
\usepackage{enumitem}
\usepackage{color}
\allowdisplaybreaks[1]
\usepackage{tikz}
 \tikzset{node distance=2cm, auto}
\usepackage{bbm} 
\usepackage{youngtab}
\usepackage{slashed}
\usepackage{mathtools}
\usepackage{wrapfig}
\usepackage{ytableau}
\usepackage{tabularray}

\def\F{\text{F}}

\def\ZZ{\mathbb{Z}} 
\def\RR{\mathbb{R}}

\def\tr{\text{tr}}

\def\tfrac#1#2{{\textstyle{\frac{#1}{#2}}}}

\def\bar{\overline}

\def\O{{\mathcal O}}

\def\g{{\gamma}}

\def\d{{\Delta}}

\def\l{{\lambda}}

\def\s{{\sigma}}

\def\Z{{\mathbb Z}}
\def\R{{\mathbb R}}

\usepackage{verbatim}   % comments out chunks of file
\usepackage[all]{xy}
\usepackage{bm}  
\def\Dslash{{\rlap{\raise 1pt \hbox{$\>/$}}D}}
\def\Pslash{{\rlap{\raise  1pt \hbox{$\>/$}}\,\partial}}
\newcommand{\be}{\begin{equation}}      
\newcommand{\ee}{\end{equation}}      
\newcommand{\bea}{\begin{eqnarray}}      
\newcommand{\eea}{\end{eqnarray}}

\def\mn#1{}
\def\ac#1{}
\def\tj#1{}

\theoremstyle{definition}

\def\ldangle{{\langle \! \langle}}
\def\rdangle{{\rangle \! \rangle}}

\begin{document}

\title{1-form symmetry versus large $N$ QCD}

\author{Aleksey Cherman, }
\emailAdd{acherman@umn.edu}
\author{Theodore Jacobson, }
 \emailAdd{jaco2585@umn.edu}
\author{Maria Neuzil}
 \emailAdd{neuzi008@umn.edu}
\affiliation{School of Physics and Astronomy, University of Minnesota, Minneapolis MN 55455, USA}

\abstract{We show that large $N$ QCD does not have an emergent  $\Z_N$ $1$-form symmetry. Our results suggest that a  symmetry-based understanding of (approximate) confinement in QCD would require some further generalization of the notion of generalized global symmetries.}

\maketitle

\section{Introduction}
\label{sec:intro}
Selection rules play a central role in quantum field theory (QFT).  A selection
rule is a statement that the vacuum expectation value of some operator or combination of operators $\mathcal O$ in a QFT vanishes, $\langle \mathcal{O} \rangle = 0$.  Typically correlation
functions of operators are impossible to evaluate exactly, so selection
rules amount to rare and valuable exact statements about a QFT.  The general
expectation is that every robust selection rule (that is, a selection rule
$\langle \mathcal{O} \rangle = 0$ which is insensitive to small parameter
variations) is associated with a global symmetry. Indeed, from a theoretical perspective, selection rules are generally viewed as Ward-Takahashi identities following from global
symmetries of a QFT, while experimentally, observations of selection rules are used to infer the existence of symmetries.

Selection rules give sharp definitions of many physical notions.  This paper
will focus on quark confinement in gauge theories like QCD. Quark confinement can be viewed as a statement about selection
rules for line operators. Specifically, confinement is associated with an area law behavior for large
contractible Wilson loops and the vanishing of Polyakov loop (non-contractible Wilson loop) expectation values.
These selection rules encode the physical ideas that widely separated
quark-anti-quark pairs experience a potential which increases linearly in their separation, and that an
isolated quark has an infinite free energy. 
Wilson loop selection rules are believed to
be associated with global symmetries that act on line operators. The first
example of this kind of symmetry was found in the late 1970's in pure gauge
theories, most notably in pure $SU(N)$ Yang-Mills (YM) theory, and dubbed `center
symmetry'~\cite{Polyakov:1978vu,Susskind:1979up,Gross:1980br,Weiss:1980rj}.  Center symmetry acts on Polyakov loops, explaining why their expectation values vanish in a confining phase.  In
recent years the notion of symmetry in quantum field theory has broadened
immensely, and center symmetry is typically viewed as a special case of a $1$-form
symmetry~\cite{Gaiotto:2014kfa}. Put simply, a $1$-form symmetry is a symmetry that acts on line operators.
The modern approach defines a global symmetry as the existence of certain
topological operators which act on charged objects and hence generate the
symmetry. In the case of $1$-form symmetries, these symmetry generators are
codimension-$2$ topological operators.  In $SU(N)$ YM theory without matter there is a $\Z_N$ $1$-form symmetry which acts on all Wilson loops, and it can be used to give a direct explanation of both the Polyakov loop and contractible Wilson loop selection rules, as we review in Appendix~\ref{sec:selectionrules}. In pure YM theory, center symmetry is just a special case of a $\Z_N$ $1$-form symmetry, and it is tempting to assume that this is a general statement.

Quark confinement is not a sharply-defined notion in real-world QCD with $N=3$
colors and $N_f = 3$  flavors of relatively light fundamental-representation
quarks, see e.g.~\cite{Greensite:2003bk} for a review.  Physically, in QCD, dynamical fundamental matter fields cause long confining strings to break, the expectation values of large contractible Wilson loops have perimeter law behavior rather than area law, and Polyakov loop expectation values are non-zero. Correspondingly, the presence of minimally-charged matter fields explicitly breaks the $\Z_N$ center and $1$-form symmetries.  

However, QCD becomes simpler in 't Hooft's large $N$ limit, where the number
of colors $N \to \infty$ while $N_f$ and $\lambda = g^2 N$ are held fixed.  Many qualitative and some quantitative features of QCD can be explained by assuming that $1/N$ corrections are fairly small for $N = 3$, see e.g.~\cite{Witten:1979kh}.  
The large $N$ limit also
allows one to give a sharp definition of confinement.  Quark loops are usually
suppressed by powers of $1/N$ at large $N$, and as a result the correlation
functions and selection rules for typical gluonic observables in large $N$ QCD coincide
with those of pure YM theory in the large $N$ limit.

This makes large $N$ QCD a confining theory:  large contractible
Wilson loops have an area law, and Polyakov loop expectation values vanish. Since it is generally expected that selection rules should be explained by symmetries, it is therefore tempting to believe that large $N$ QCD should have all of the symmetries of
large $N$ pure YM theory, so that both the $\Z_N$ center symmetry and $\Z_N$
$1$-form symmetry of YM theory survive the introduction of finite-mass
fundamental-representation quarks. For this to be the case, large $N$ QCD should have codimension-$2$ operators which are topological up to
$1/N$ corrections, just as in the case of $1$-form symmetries that emerge due to
large mass limits, where there are codimension-$2$ operators which are topological on distances large compared to the inverse quark mass~\cite{Cordova:2022rer}. 

%%%%%%%%%%
\begin{figure}[h!] \centering \includegraphics[width=0.4\textwidth]{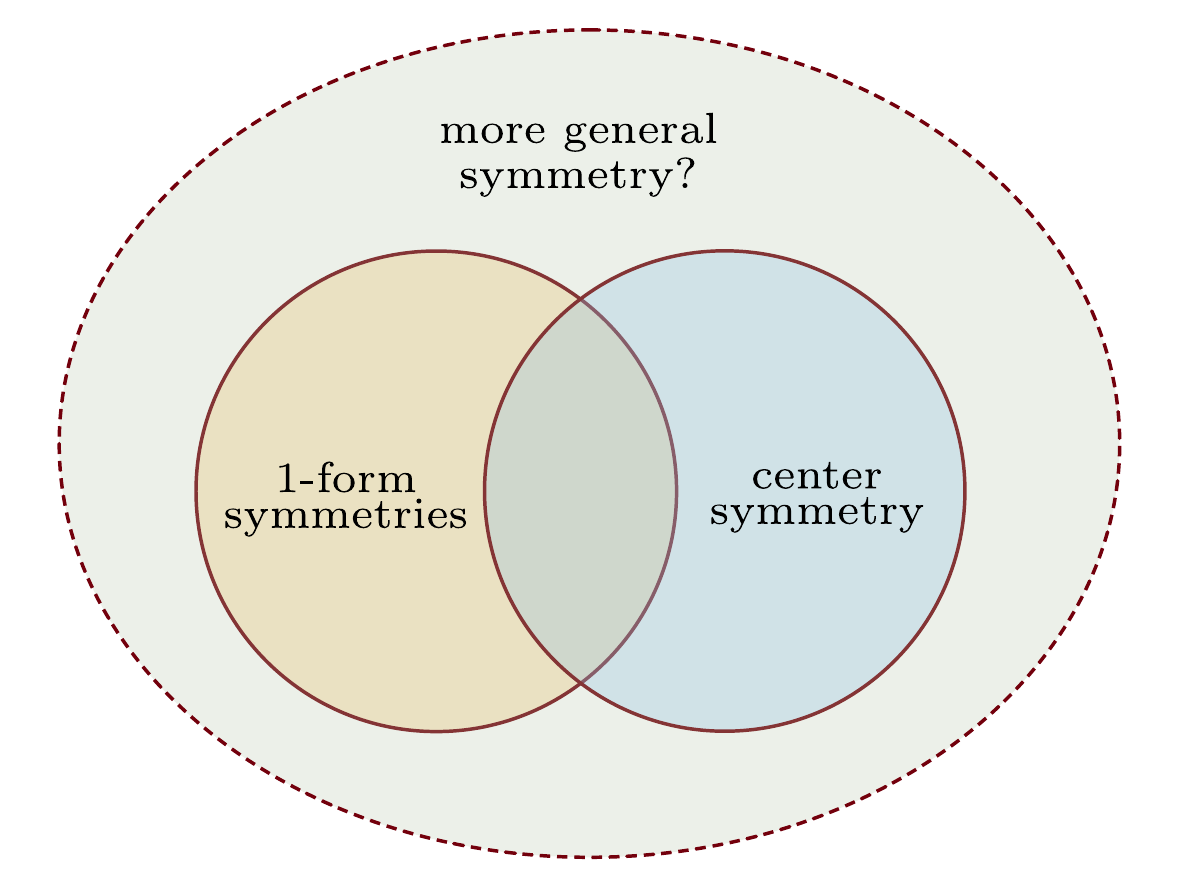}
\caption{Wilson loops in large $N$ QCD with quarks of mass $m$ obey the same selection rules as in large $N$ pure YM theory.  Some of these selection rules (the ones inside the blue circle) can be explained by a $\Z_N$ center symmetry.  Some of them can be explained by a $\Z_N$ $1$-form symmetry (beige circle), but only in the YM limit $m \to \infty$. There is no $\Z_N$ $1$-form symmetry in large $N$ QCD with finite $m$, and so there are selection rules outside both the blue and beige circles. Moreover, as we review in Section~\ref{sec:conclusions}, there are also many large $N$ selection rules that cannot be explained by a $\Z_N$ $1$-form symmetry even when $m \to \infty$.} \label{fig:center_1_form_venn_new} 
\end{figure}
%%%%%%%%%%%%%%

In this paper we show that only one of the expectations above is satisfied. Large
$N$ QCD does have a $\Z_N$ center symmetry.  However, it does not have a $\Z_N$ $1$-form
symmetry.\footnote{Here we focus on the large $N$ limit with quarks in the fundamental representation of $SU(N)$.  At $N=3$ the (anti-)fundamental representation is isomorphic to the two-index anti-symmetric (AS) representation, so there is also an interesting `QCD[AS]' large N limit where the quarks are in the AS representation and $N_f$ is again fixed, see e.g. the very end of~\cite{tHooft:1973alw}, and also~\cite{Armoni:2004uu} for an early review of an extensive literature triggered by~\cite{Armoni:2003gp}.  Armoni, Shifman and \"Unsal~\cite{Armoni:2007kd} have also argued that QCD[AS] has an emergent $\Z_N$ center symmetry at large $N$ despite the non-suppression of quark loops for any $N$.  The same reference emphasized that QCD[AS] has stable $k$-strings, as does its orientifold-equivalent partner, QCD[adj]. However, strictly speaking, from the more recent generalized symmetry perspective, a symmetry explanation of the stability of $k$-strings would require a $\Z_N$ $1$-form symmetry, not just a $\Z_N$ center symmetry.  Our arguments in this paper against the emergence of a $\Z_N$ $1$-form symmetry at large $N$ also apply to QCD[AS].}  This implies that at least in large $N$ QCD, center symmetry is not simply a special case of  $1$-form symmetry.  
These notions are distinct; see Fig.~\ref{fig:center_1_form_venn_new} for an illustration.

To demonstrate our conclusions, we start by reviewing the notion of $1$-form symmetry in Section~\ref{sec:one_form_YM}, and discuss center symmetry and its relation to $1$-form symmetry in Section~\ref{sec:center_sym}.  We then discuss the obstructions to the emergence of a $1$-form symmetry in large $N$ QCD in Section~\ref{sec:obstructions}.  First, we explain a general obstruction 
to the emergence of $\Z_N$ $1$-form symmetry due to the
Rudelius-Shao notion of endability~\cite{Rudelius:2020orz} in Section~\ref{sec:endability}. An implication of the endability obstruction is that quark loop effects are not suppressed in correlation functions of the
would-be topological operators that naively should generate a $1$-form symmetry. We give an independent argument for this non-suppression of quark loops in Section~\ref{sec:quark_loop}.
In Section~\ref{sec:2dQCD}, we show that these general arguments are consistent with the results of
explicit calculations in 2d QCD performed using a hopping parameter expansion on the lattice.  Finally, in Section~\ref{sec:conclusions} we summarize and contextualize our observations, and suggest some directions for future work.  Appendix~\ref{sec:selectionrules} contains a pedagogical discussion of the derivation of selection rules from $1$-form symmetries, covering both invertible and non-invertible symmetries.  We also comment on the derivation of selection rules from non-invertible $0$-form symmetries. Appendix~\ref{sec:2dYM} is a review of 2d pure YM on the lattice.  Appendix~\ref{sec:vortex_correlator_appendix} contains some results on the correlation functions of Wilson lines and center vortices (the would-be $1$-form symmetry generators) which supplement the discussion in the main text.

%%%%%%%%%%%%%%%%%%
\section{$\Z_N$ $1$-form symmetry in pure YM theory}
\label{sec:one_form_YM}
%%%%%%%%%%%%%%%%%%

\begin{figure}[h]
\centering
\includegraphics[width=0.8\textwidth]{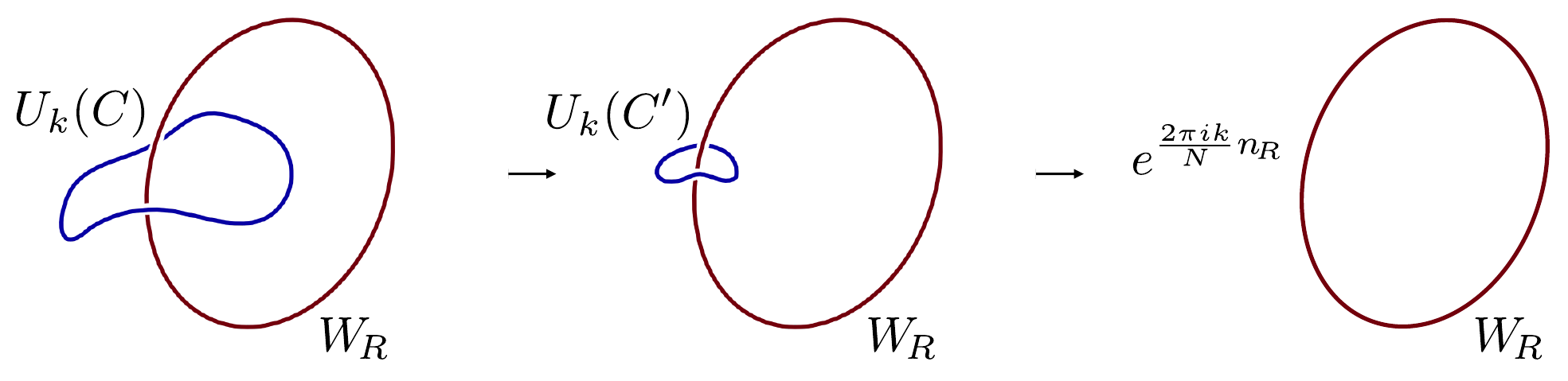}
\caption{Action of a $\Z_N$ $1$-form symmetry generator $U_k(C)$ on a Wilson loop in representation $R$ in $d=3$.}
\label{fig:Wilson_loop_action}
\end{figure}

To set the stage, in this section we briefly review the $1$-form symmetry of
$SU(N)$ gauge theory without matter (pure YM theory).   This $\Z_N$ symmetry is defined in terms of its generators $U_k$, which are topological operators
supported on $(d-2)$-dimensional closed manifolds $\Sigma_{d-2}$,
\begin{align}
    U_{k}(\Sigma_{d-2}) \,.
\end{align}
These operators have only a topological dependence on $\Sigma_{d-2}$, and satisfy a
$\mathbb{Z}_N$ `fusion rule'
\begin{align}
    U_{k}(\Sigma_{d-2}) U_{\ell}(\Sigma_{d-2})  =  U_{k+\ell \textrm{ mod } N}(\Sigma_{d-2}) \,.
    \label{eq:ZN_fusion}
\end{align}
In $d>2$, $\langle U_{k}(\Sigma_{d-2})\rangle = 1$ when $\Sigma_{d-2}$ is
contractible, while in $d=2$ where $\Sigma_{d-2}$ is a point,  $\langle
U_{k}(\Sigma_{d-2})\rangle = e^{\frac{2\pi i k}{N}\ell},\, \ell = 0,1,\ldots N-1$. In
the presence of a representation $R$ Wilson loop on a closed simple contour $C$,
these operators satisfy the $d>2$ relation\footnote{There is a closely-related
relation in $d=2$ which we discuss later.}
\begin{align}
    U_{k}(\Sigma_{d-2}) W_{\! R}(C) = e^{\frac{2\pi i k}{N}n_{ R}}\, W_{\! R}(C) 
\end{align}
where we have assumed that $\Sigma_{d-2}$ and $C$ have unit linking number on
the left hand side, $n_{R}$ is the $N$-ality of $R$,\footnote{The $N$-ality
of an irreducible representation of $SU(N)$ is equal to the number of boxes mod
$N$ in the associated Young diagram.
% , or more generally the charge under the $\ZZ_N$ center of the group.
} and to pass from the
left side to the right side $\Sigma_{d-2}$ is shrunk to a point while preserving
the linking number, as illustrated in Fig.~\ref{fig:Wilson_loop_action} for
$d=3$.

%%%%%%%%%%
%%%%%%%%%%%%%%

Now let us consider the large $N$ limit.  The Wilson loops which are most
relevant to the large $N$ theory are those with $N$-ality $n_{R} =
\mathcal{O}(1)$.  Representations with $N$-alities of $\O(N)$ are expected to
have string tensions scaling as a positive power of $N$, and hence have
expectation values which vanish exponentially with $N$, see
e.g.~\cite{ARMONI200367}. At large $N$, the action of the minimal 1-form
symmetry operator with $k = 1$ on a Wilson loop with $n_{ R} = \mathcal{O}(1)$
is
\begin{align} 
\label{eq:trivialphase}
    U_{1}(\Sigma_{d-2}) W_{\! R}(C) = e^{\frac{2\pi i }{N}n_{ R}}\, W_{\! R}(C) = W_{\! R}(C) + \O\left(\frac{1}{N}\right)\,.
\end{align}
This means that $U_k(\Sigma_{d-2})$ with $k = \mathcal{O}(1)$ act trivially on the operators with non-vanishing expectation values at large $N$, and in this sense $U_k(\Sigma_{d-2})$ with $k = \mathcal{O}(1)$ become trivial operators at large $N$.  Trivial operators should have unit expectation values, and we will see in Section~\ref{sec:2dQCD} that this is indeed true for $U_k(\Sigma_{d-2})$ with $k = \mathcal{O}(1)$, up to $1/N$ corrections.  

The triviality of $U_k(\Sigma_{d-2})$ with $k = \mathcal{O}(1)$ at large $N$ means that these generators cannot be used to derive selection rules for e.g. fundamental Wilson loops in large $N$ YM theory. To get a phase in Eq.~\eqref{eq:trivialphase} which does not vanish in the large $N$ limit, one must take $\min (k,N-k)  \sim N$. Therefore, a natural criterion for a theory to have a non-trivial 1-form symmetry at large $N$ is that at least the subset of the operators $U_k(\Sigma_{d-2})$ with $k = \O(N)$ must be topological.  This criterion is of course satisfied by pure YM theory, where the operators  $U_k(\Sigma_{d-2})$ are topological for all $k$.

We should also comment on the expectation values of $U_{k}(\Sigma_{d-2}) $ operators.  When the spacetime manifold is $\mathbb{R}^d$, the cluster decomposition principle along with the $\Z_N$ fusion rules obeyed by these operators imply that
\begin{align}
    \langle U_{k}(\Sigma_{d-2})  \rangle = e^{\frac{2 \pi i k}{N}  \ell} 
    \,.
    \label{eq:VEV_constraint}
\end{align}
where $\ell$ is an integer (defined mod $N$). If we multiply the right-hand side by a number $c_k \in \mathbb{R}_{+}$, the $\mathbb{Z}_N$ fusion rule will be violated unless $c_k c_n = c_{k+n \text{ mod }N}$, which is impossible for positive $c_k$ unless it is equal to $1$ (or $0$).  In $d>2$, one can also argue that $\ell = 0$ for contractible $\Sigma_{d-2}$, while in $d = 2$ the value of $\ell$ in the vacuum state is a dynamical question.

%%%%%%%%%%%%%%
\section{$\Z_N$ center symmetry}
\label{sec:center_sym}
%%%%%%%%%%%%%%

We now review the long-known argument (see e.g. Ref.~\cite{Armoni:2007kd}) that large $N$
QCD has a center symmetry when placed on spacetime manifolds of the form $M_{d-1}
\times S^1$. The standard boundary conditions on $S^1$ for the 
gauge fields $a_{\mu}$, quark fields $q$, and gauge transformations $g\in SU(N)$ are
\begin{align}
\begin{split}
a_{\mu}(\vec{x}, x_4 +\beta) &= a_{\mu}(\vec{x},x_4)\\ 
q(\vec{x}, x_4 +\beta) &= - q(\vec{x},x_4)\\ 
g(\vec{x},x_4+\beta) &= g(\vec{x},x_4) \,,
\end{split}
\end{align}
where we have (arbitrarily) chosen `thermal' boundary conditions for the quarks.  A center symmetry transformation is defined in terms of gauge equivalence
classes of functions  $[h_k] = \{ h_k(x^{\mu}) \in SU(N),
h_k(x^{\mu}) \sim g(x^{\mu}) h_k(x^{\mu}) g(x^{\mu})^{\dag} \}$ which are periodic up to the center,
$h_k(x_4 = \beta) = e^{\frac{2\pi i k}{N}}h_k(x_4 =0)$.  We can write the center symmetry
transformation by picking a representative $h_k(x^{\mu})$ of one of the equivalence classes and transforming
\begin{align}
     a_\mu(x) \to h_k(x)a_\mu(x)h_k^\dagger(x) + i h_k(x)\partial_\mu h_k^\dagger(x).
     \label{eq:center_transform}
\end{align}
%Note that \eqref{eq:center_transform} is \emph{not} a gauge transformation. It
%is not periodic, and it does not act on the quarks.  
In the pure YM limit where the quark mass $m \to \infty$, this is clearly a
symmetry of the path integral weight $e^{-S_{\text{YM}}[a_{\mu}]}$.  Center
transformations do not act on any local operators or on contractible Wilson loops. However, center transformations do act on Polyakov
loops,
\begin{align}
    \frac{1}{d_R} \tr_{R} P(\vec{x},z)   \to e^{\frac{2\pi i k }{N}n_R}   \frac{1}{d_R} \tr_{ R} P(\vec{x},z) 
    \label{eq:center_action_Polyakov}
\end{align}
where $P = \mathcal{P} e^{i \oint_{S^1} a_4}$.  This means that e.g. $\langle
\frac{1}{N} \tr_{\F} P(\vec{x})\rangle $ is an order parameter for center
symmetry.  In the `t Hooft large $N$ limit of QCD the quark part of the action
is $1/N$ suppressed relative to the pure glue part of the action, and so
\eqref{eq:center_transform} is also a symmetry of QCD at large $N$.   Center
symmetry in large $N$ QCD predicts that as external parameters such as
temperature are varied, the Polyakov loop expectation values should either
vanish up to $1/N$ corrections (indicating an approximate $\Z_N$ center symmetry
which is not spontaneously broken), or there should be $N$ nearly-degenerate
states with different phases for the Polyakov loop and splittings suppressed by
$1/N$ (indicating a spontaneously broken approximate center symmetry).  These
predictions can be verified by e.g. calculations with appropriate
Gross-Pisarski-Yaffe effective
potentials~\cite{Gross:1980br,Myers:2009df,Unsal:2010qh}, and so large $N$ QCD
clearly has a $\Z_N$ center symmetry, just as pure YM theory does.

%%%%%%%%%%%%%%%%%%
\subsection{Relation between center symmetry and $1$-form symmetry in pure YM}
\label{sec:center_vs_one_form}
%%%%%%%%%%%%%%%%%%

The existence of a $\Z_N$ $1$-form symmetry implies the existence of a $\Z_N$
center symmetry. To see why, we first recall  that the physical content of a
symmetry is in how it acts on correlation functions rather than in how it acts
on fields in a particular path integral presentation of the theory. The action
of center symmetry on correlation functions is given in
Eq.~\eqref{eq:center_action_Polyakov}.  As we explain in detail in
Appendix~\ref{sec:selectionrules}, in a theory with a $1$-form symmetry this
transformation of Polyakov loops in finite volume can be deduced from the
correlation function
\begin{align}
    \left\langle U_{-k}(M_2) U_k(M_2')\, \tr_{R} P \right\rangle \,.
    \label{eq:center_from_one_form_corr}
\end{align}
The topological nature of the $U_k$ operators implies that we can deform and
move $M_2'$ so that it merges with $M_2$ without crossing the Polyakov loop,
which means that this correlator is equal to  $\left\langle \tr_{R} P
\right\rangle$.  Alternatively, we can deform and move $M'_2$ so that it merges
with $M_2$ having crossed the Polyakov loop once, from which we see that
\eqref{eq:center_from_one_form_corr} is also equal to $e^{\frac{2\pi i k}{N}n_R}
\left\langle \tr_{R} P \right\rangle$.   This means that Polyakov loops
transform in precisely the way implied by center symmetry
\eqref{eq:center_action_Polyakov}.  An entertaining implication of this argument
is that a $\Z_N$ center transformation in a theory with a $1$-form $\Z_N$ symmetry can be
thought of as insertion of the identity operator, which is then resolved into a pair of nearby
codimension-$2$ topological operators, $1 \sim U_{-k}(M_2) U_k(M_2')$.  

The fact that large $N$ QCD has a $\Z_N$ center symmetry and also obeys the same
selection rules for contractible Wilson loops as pure YM theory makes it
tempting to guess that it should have a $1$-form symmetry.  However, this
requires that it have codimension-$2$ topological operators with the right
properties.  In the next sections we explain obstructions to the existence of
such operators in large $N$ QCD.  These obstructions mean that the existence of
a $\Z_N$ center symmetry does not imply the existence of a $\Z_N$ $1$-form
symmetry.

%%%%%%%%%%%
\section{Obstructions to $1$-form symmetry in large $N$ QCD}
\label{sec:obstructions}
%%%%%%%%%%%

In this section we explain two general obstructions to the existence of a $\Z_N$
$1$-form symmetry in QCD in the large $N$ limit.  One obstruction is due to the
endability of Wilson lines in QCD, while the other is due to non-suppression of
quark loops in some correlation functions in large $N$ QCD.

%%%%%%%%%%%%
\subsection{Endability}
\label{sec:endability}
%%%%%%%%%%%%

\begin{figure}[h]
\centering
\includegraphics[width=0.9\textwidth]{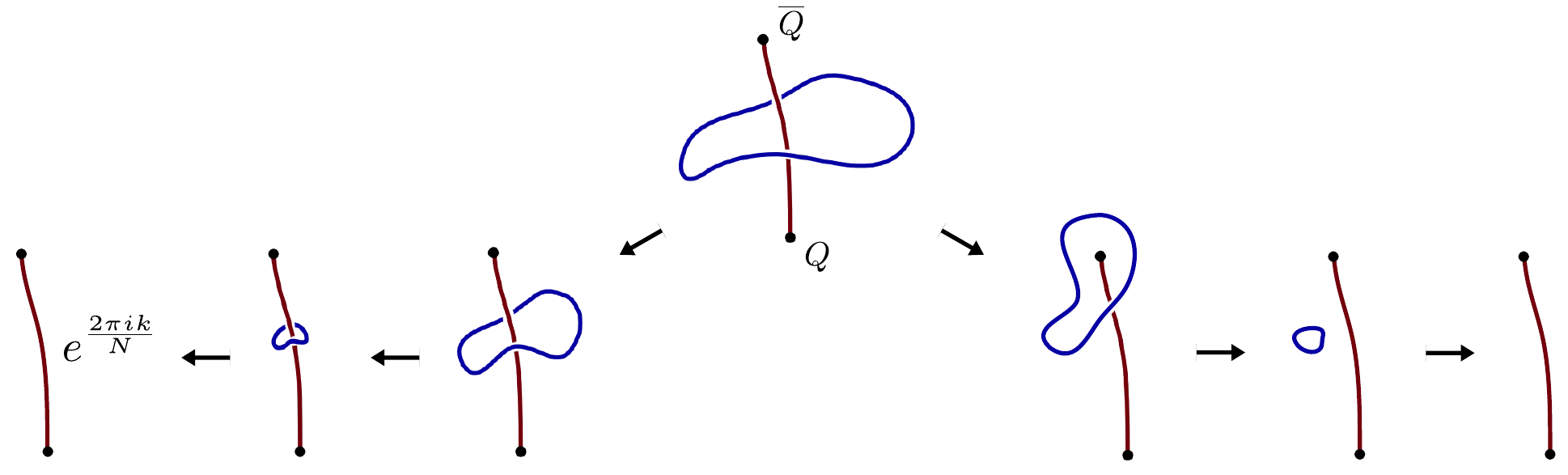}
\caption{Different ways of shrinking a topological codimension-2 operator in the presence of an open fundamental Wilson line give different results. This gives rise to the so-called `endability' constraints on the existence of topological operators with non-trivial actions on Wilson loops. Here we depict the situation in three spacetime dimensions, where the would-be topological line can be collapsed in two distinct ways.}
\label{fig:Wilson_line_action}
\end{figure}

The appearance of a $\Z_N$ $1$-form symmetry in the large $N$ limit of QCD should be tied to the existence of codimension-$2$ operators $U_k(\Sigma_{d-2})$ with the properties discussed in Section~\ref{sec:one_form_YM}. These operators ought to be only approximately topological for large but finite $N$, unless we also send $m/\Lambda \to \infty$. The notion of `endability' introduced in the work of Rudelius and Shao~\cite{Rudelius:2020orz} yields an obstruction to this scenario.  When a gauge theory has a dynamical fundamental matter field $Q$, Wilson lines can end on $Q$.  As a result, large $N$ QCD has gauge-invariant open Wilson line operators, such as
\begin{align}
    W_{\F}(C_{x,y}) = \frac{1}{N} \overline{Q}(x) \mathcal{P} \exp\left({i \int_{C_{x,y}}\!a}\right) Q(y)\,,
    \label{eq:open_line}
\end{align}
where $C_{x,y}$ is a curve connecting the points $x$ and $y$ and we have assumed that $Q$ is a Dirac fermion to be concrete. 
When $W_{\F}(C_{x,y})$ is an allowed gauge-invariant operator, there cannot be non-trivial topological codimension-$2$ operators which act on Wilson lines~\cite{Rudelius:2020orz}.  
To see why, consider Fig.~\ref{fig:Wilson_line_action}, and suppose there exists a topological operator $U_k(\Sigma_{d-2})$. (The figure depicts the situation in $d=3$, where the would-be topological operator is supported on a closed curve.) 
Then one could shrink $U_k(\Sigma_{d-2})$ around the open Wilson line yielding $e^{2\pi i k/N}  W_{\F}(C_{x,y})$.  Alternatively, one could use the assumed topological property of $U_k(\Sigma_{d-2})$ to move it past the endpoint of the Wilson line, and then shrink it, yielding just $W_{\F}(C_{x,y})$.  
This is a contradiction, because the correlation function was assumed not to change as we move the would-be topological operator $U_k(\Sigma_{d-2})$. Therefore $U_k(\Sigma_{d-2})$ cannot be a topological operator. 

%%%%%%%%%%%%%
%%%%%%%%%%%%%%

How can one escape this conclusion? First, we note that a valid way to evade the constraints from endability is if $\langle W_{\F}(C_{x,y}) \rangle \to 0$ in the limit where one expects the operators $U_k(\Sigma_{d-2})$ to become topological.\footnote{In fact, one would have to require that in the limit of interest, the open line acts as zero in all correlation functions. This stronger condition is obtained by repeating the argument depicted in Fig.~\ref{fig:Wilson_line_action} in the presence of additional operator insertions separated from the open Wilson line. \interfootnotelinepenalty=10000}  This is precisely what happens in the large mass limit $m/\Lambda \to \infty$, and so in this pure-YM limit one can expect an approximate $1$-form symmetry to be generated by approximately-topological codimension-$2$ operators; see Ref.~\cite{Cordova:2022rer} for a detailed discussion of the related situation with $U(1)$ $1$-form symmetry.  
However, this escape is not available in large $N$ QCD with finite $m/\Lambda_{QCD}$, because $\langle W_{\F}(C_{x,y}) \rangle \neq 0$ in the large $N$ limit with fixed quark mass $m$.  One quick way to check this is to note that in the limit that $C_{x,y}$ is short, $\langle W_{\F}(C_{x,y}) \rangle$ approaches the expectation value of a chiral condensate in massive 4d QCD, normalized such that it scales as $\sim \Lambda^3$ in terms of the strong scale $\Lambda$ when e.g. $m \sim \Lambda$. As we discuss in detail in Section~\ref{sec:2dQCD}, this $N$-scaling can also be verified using hopping and strong-coupling expansions on the lattice.

Accepting the existence of nontrivial open Wilson lines in large $N$ QCD, one might still question the extent to which the endability argument constrains the behavior of would-be symmetry operators in correlation functions with only \emph{closed} Wilson lines. Indeed, one could attempt to use the fact that correlation functions of $\mathcal{O}(1)$ single-trace operators factorize at large $N$, so that
\begin{align}
    \langle W_{\F}(C) W_{\F}(C_{x,y}) \rangle = \langle W_{\F}(C) \rangle \langle  W_{\F}(C_{x,y}) \rangle + \mathcal{O}\left(\frac{1}{N}\right) \,.
\end{align}
One can therefore (correctly) argue that the closed Wilson loop operators define a consistent subsector of the theory. One might go further and try to argue that in large $N$ QCD, the operators $U_k(\Sigma_{d-2})$ are topological in correlation functions which only contain closed Wilson loops, so that for example
\begin{align}
\langle U_k(\Sigma_{d-2}) W_{\! R }(C)\rangle \stackrel{?}{=}
e^{\frac{2\pi i k  }{N}n_{ R}\, \ell(\Sigma_{d-2}, C)}\, \langle  W_{\! R }(C)\rangle\,,
\label{eq:Z_N_question_QCD}
\end{align}
and $\ell$ is the linking number. However, such an argument is problematic. First, this reasoning contradicts the expectation that the action of $U_k(\Sigma_{d-2})$ on $W_{\! R }(C)$ should be determined from a \emph{local} computation in some small region near $C$. This means that there is no way for $U_k(\Sigma_{d-2})$ to `know' whether the Wilson loop it is trying to act on is open or closed.  So if $U_k(\Sigma_{d-2})$ has a non-topological action on open Wilson lines, then it must also have a non-topological action on closed Wilson lines, and the correlation functions of the $U_k(\Sigma_{d-2})$ operators themselves should not be topological. 

Similarly, the above argument assumes that correlation functions that involve $W(C)$ and $U_k(\Sigma_{d-2})$ in large $N$ QCD coincide with the values of the analogous correlation functions between Wilson loops and $\Z_N$ $1$-form symmetry generators in pure YM theory. The reason one might think this should be true is that quark loops are suppressed at large $N$ in typical gluonic observables. However, we will see in Section~\ref{sec:quark_loop} below that quark loop effects are in general \emph{not} suppressed in correlation functions that involve $U_k(\Sigma_{d-2})$.

Finally, this sort of imagined `subsector 1-form symmetry' would not be powerful enough to directly explain many selection rules in large $N$ QCD such as 
\begin{align}
    \left\langle \frac{1}{N} \tr_{\F} P(\vec{x}) \,\, \bar{Q}Q(x^{\mu})  \right\rangle = 0 \,,
\end{align}
because near the continuum limit we can replace $ \bar{Q}Q(x^{\mu})$ by its slightly-smeared version $W_{\F}(C_{x^{\mu},x^{\mu} + \epsilon^{\mu}})$, and then the insertions of $U_k(\Sigma_{d-2}) U^{\dag}_k(\Sigma'_{d-2})$ which are needed to prove the selection rule (see Appendix~\ref{sec:selectionrules}) fail to be topological.

% More concretely, while quark loop contributions are indeed suppressed in many well-studied observables in large $N$ QCD, as in Eq.~\eqref{eq:largeNWilsonRelation}, one can verify by explicit calculation that their effects are \emph{not} $n_f/N$ suppressed in correlation functions that involve $U_k(\Sigma_{d-2})$.  We present such a calculation in Section~\ref{sec:2dQCD}.

%%%%%%%%%%%%
\subsection{Quark loop non-suppression}
\label{sec:quark_loop}
%%%%%%%%%%%%
We now address the issue of quark loop suppression.  The general lore is that dynamical quark loop effects are $1/N$ suppressed in large $N$ QCD. If this standard lore is correct then we expect that the correlation functions of would-be $1$-form symmetry generators like $\langle U_k(\Sigma_{d-2})\rangle$ and $\langle U_k(\Sigma_{d-2}) W_{\! R }(C)\rangle$ in large $N$ QCD would be the same as in pure YM theory.  The endability argument we gave above implies that this conventional lore must be wrong in the present situation. 

To see how this is possible, suppose that quark loop effects are suppressed in correlation functions of e.g. $U_1(\Sigma_{d-2})$. To write concrete expressions we will take $d=2$ in the rest of this subsection, but the arguments easily generalize to $d>2$. The structure of the $1/N$ expansion in large $N$ QCD makes it natural to expect that the vacuum expectation value of $U_1(x)$ and its two-point correlation function satisfy
\begin{subequations}
\label{eq:U1_correlators}
\begin{align}
    &\langle U_1(x)\rangle = 1 + v/N\\
    &\langle U_1^{\dag}(x) U_1(0) \rangle = 1 + u(x)/N \\
    &\langle U_1(x) U_1(0) \rangle = 1 + w(x)/N
\end{align}
\end{subequations}
where $v, u,w \sim \mathcal{O}(1)$, $u(x), w(x)$ are some decreasing functions of $x$, and we have assumed that in the vacuum state the parameter $\ell$ from Eq.~\eqref{eq:VEV_constraint} vanishes, as would be the case in 2d pure YM theory in the large volume limit. However, as we discussed around Eq.~\eqref{eq:trivialphase}, the operators $U_k(x)$ with $k = \mathcal{O}(1)$ are trivial operators up to $1/N$ corrections in the large $N$ limit. The crucial question is therefore whether we should expect quark loop effects to be suppressed in correlation functions that involve the non-trivial would-be generators of the $1$-form symmetry $U_k(x)$ with $k = \O(N)$.

We will argue by contradiction. Suppose that the $U_k$ operators are topological at large $N$ for all $k$ in large $N$ QCD, which would mean that quark loop effects are suppressed for all $k$, not just for $k = \mathcal{O}(1)$.  Then the expectation values must satisfy Eq.~\eqref{eq:VEV_constraint}, so that $|\langle U_k(x) \rangle| = 1$. Next, the assumption that the $U_k$ operators are topological means that we can think of $U_k(x)$ as $k$ insertions of the $U_1$ operator at slightly separated points:
\begin{align}
    \langle U_k(x)\rangle = \langle U_1(x+\epsilon_1)U_1(x+\epsilon_2) \cdots U_1(x+\epsilon_k)\rangle
    \label{eq:Uk_split}
\end{align}
If the $U_1$ operators were \emph{exactly} topological, the connected parts of this correlator would all vanish, and we would get $\langle U_k(x)\rangle = \langle U_1(x)\rangle^{k} = 1$.  This is what happens in pure YM theory. But in large $N$ QCD, Eq.~\eqref{eq:U1_correlators} leads to $\langle U_k(x)\rangle \sim \langle U_1(x)\rangle^{k} \sim \left(1+v/N\right)^{k}$.  If $k = \mathcal{O}(1)$, our assumptions do not lead to any contradictions. But  if $k = \bar{k} N$, then our assumptions lead us to  $\langle U_k(x)\rangle \sim e^{v \bar{k}}$ as $N\to\infty$.  This deviates from unity if $\bar{k} = \O(1)$, which is a contradiction.  So quark loop suppression must fail for $k = \O(N)$, and we should not expect $U_k(x)$ to be topological for all $k$.
 
In fact, the breakdown of quark loop suppression actually should set in even before $k = \O(N)$, kicking in no later than $k \gtrsim \O(\sqrt{N})$.  To see this we consider the correction to factorization of Eq.~\eqref{eq:Uk_split} into $k$ VEVs.  We can estimate the magnitude of this correction from the connected parts of the two-point functions in Eq.~\eqref{eq:U1_correlators}. There are $\sim k^2$ ways to choose two operators from $k$ operators, so we can estimate the non-completely-factorized part of Eq.~\eqref{eq:Uk_split} as $\sim k^2/N$, which becomes $\mathcal{O}(1)$ once $k = \O(\sqrt{N})$.

The breakdown of quark loop suppression discussed above is similar to the breakdown of large $N$ factorization for correlation functions of $k$ single-trace operators in pure YM theory for $k \sim N$.  This makes the non-emergence of $\Z_N$ $1$-form symmetry in large $N$ QCD indicated by the endability argument of Section~\ref{sec:endability} consistent with the expected relations between correlation functions of gluonic operators in large $N$ QCD and large $N$ YM theory. 

%%%%%%%%%%%%
\section{Scalar QCD in two dimensions}
\label{sec:2dQCD}

To gain more insight into the discussion above, in this section we focus on one of the simplest theories where a 1-form symmetry might have been expected to emerge at large $N$, namely two-dimensional $SU(N)$ gauge theory coupled to a single fundamental scalar $\phi$. We take the Euclidean spacetime to be a torus, and regularize the theory on a square lattice with sites $x$, links $\ell$, and plaquettes $p$.  We choose units such that the lattice spacing $a = 1$.  We take $u_\ell \in SU(N)$ to be group-valued link variables, $\phi_x$ to be the scalar fields, and $u_p$ will will denote the  (path-ordered) product of link variables around the boundary of a plaquette, $u_p =\mathcal P \prod_{\ell\in\partial p}u_\ell$.   The partition function is given by
\begin{align} \label{eq:2dqcdZ} 
Z = \prod_\ell \int du_\ell \prod_x \int d\phi_x d\phi^\dagger_x\, \prod_p e^{-s_{\text{YM}}(u_p)} \prod_\ell e^{-s_{\text{H}}(\phi_x^\dagger u_\ell \phi_{x'} )} \prod_x e^{-m^2 \phi_x^\dagger \phi_x}\,,
\end{align}
where $\int du_\ell$ is the Haar measure and for later convenience we define $\int d\phi_x = \prod_{i=1}^N \sqrt{\tfrac{m^2}{\pi}}\int d\phi_{x,i}$, where $i$ is a color-fundamental index. The hopping terms which couple the gauge and matter fields take the form
\begin{equation}
s_{\text{H}}(\phi_x^\dagger u_\ell \phi_{x'}) = - \kappa\, \phi_x^\dagger u_\ell \phi_{x'} + \text{h.c.}\,,
\end{equation}  
where $\kappa$ is the hopping parameter, and $m$ is the mass parameter.   The most familiar gauge action is the Wilson action
\begin{equation}
    s_{\text{YM}}(u_p) = s_{\text{Wilson}}(u_p) = -\frac{1}{2g^2}\tr(u_p + u_p^\dagger) \,.
\end{equation}
Analytic calculations with this action use the strong-coupling expansion in the parameter $1/g^2$.  Strong coupling expansions and hopping expansions in the parameter $\kappa$ (which are really large mass expansions in the parameter $\kappa/m^2$) are known to have a finite radius of convergence at least at fixed $N$, including for suitably localized observables in the thermodynamic limit; see e.g.~\cite{Osterwalder:1977pc,Yaffe:1981vf, Seiler_book}.

Since we are in two dimensions, we find it useful to instead use the subdivision-invariant heat-kernel action~\cite{Migdal:1975zg,Drouffe:1978py,Lang:1981rj,Menotti:1981ry,Witten:1991we}
\begin{equation}
e^{-s_\text{YM}(u_p)} = \sum_\alpha d_\alpha \chi_\alpha(u_p)\, e^{-g^2 c_\alpha  A_p}\,,
\end{equation}
where the sum runs over irreducible representations $\alpha$ with trace $\chi_\alpha$, $A_p = a^2 = 1$ is the area of a single plaquette, and the dimension and quadratic Casimirs of the representations are denoted $d_\alpha = \chi_\alpha(\mathbbm{1})$ and $c_\alpha$, respectively. The heat-kernel action has the nice conceptual feature that it automatically yields continuum-limit answers for the pure gauge theory, and it can be viewed as a Wilson action supplemented by an infinite number of cunningly-chosen improvement terms, see e.g.~\cite{Menotti:1981ry}. Indeed, one can express the heat-kernel action in a more conventional from by performing a character decomposition in the exponent of the single-plaquette Boltzmann weight, 
\begin{equation} \label{eq:heatkernelweight}
\sum_\alpha d_\alpha \chi_\alpha(u_p)\, e^{-g^2 c_\alpha } = \exp\left[\sum_{\alpha}\text{Re}\frac{1}{g_\alpha^2}\chi_\alpha(u_p) \right]\,.
\end{equation}
The heat-kernel action also has the nice practical feature that it makes strong-coupling and hopping expansion calculations especially straightforward. Moreover, by working with the heat-kernel action, we can avoid the need to expand in $1/g^2$ in concrete calculations, since in 2d the heat-kernel action allows one to work exactly in $g^2$, order by order in a hopping expansion.

Pure $SU(N)$ YM theory in $d$ spacetime dimensions has an exact $\ZZ_N$ 1-form symmetry generated by topological operators supported on codimension-2 submanifolds $\tilde \Sigma_{d-2}$ on the dual lattice. There are several equivalent ways to define these operators.  The definition we find most convenient here views $U(\tilde{\Sigma}_{d-2})$ as `thin center-vortex' operators, see e.g. Refs~\cite{Bachas:1982ep,Engelhardt:1999xw,Greensite:2016pfc,Greensite:2003bk}, which are obtained by modifying the Boltzmann weights for plaquettes pierced by $\tilde \Sigma_{d-2}$.  In $d=2$, $\tilde{\Sigma}_{d-2}$ is just a point $\tilde{x}$, and the generators of the $\Z_N$ $1$-form symmetry of 2d pure YM can be written as 
\begin{equation}
U_k(\tilde x = \star p) = \exp\left[ s_{\text{YM}}(u_p) - s_{\text{YM}}(\omega^{-k} u_p)\right]\,,
\label{eq:2dcentervortex}
\end{equation}
where the site $\tilde x$ is dual to the plaquette $p$ and $\omega = e^{2\pi i/N}$.  This operator is topological essentially because the point $\tilde{x}$ can be moved by field redefinitions on the link fields.  One can verify that it satisfies the $\Z_N$ fusion rule Eq.~\eqref{eq:ZN_fusion}, and acts on Wilson loops in the expected way (see Appendix~\ref{sec:2dYM} for details):
\begin{align}
    \langle U(\tilde{x}) W_{\! R}(C) \rangle_0 
     = e^{\frac{2\pi i k}{N}n_R} \langle U(\tilde{x}') W_{\! R}(C) \rangle_0 
\end{align}
where $\tilde{x}, \tilde{x}'$ are inside and outside the closed curve $C$, respectively, and $\langle \cdot \rangle_0$ denotes an expectation value in pure YM theory. 

In principle, there are many inequivalent choices of codimension-$2$ operators in QCD which reduce to the $1$-form symmetry generators in pure YM in the limit $m\to \infty$. Any choice of codimension-$2$ operator will be subject to the general constraints laid out in Section~\ref{sec:obstructions}. Apart from a possible normalization factor which we discuss below in Section~\ref{sec:nonsuppression}, in the following we take the pure YM center-vortex operators \eqref{eq:2dcentervortex} to be the candidates for topological operators associated with the possible $1$-form symmetry in large $N$ QCD.   

In our discussion below it will be important to make use of some general constraints on the correlation functions of $U_{k}(\tilde{x})$ operators which hold regardless of whether they are topological.  To derive these constraints we note that an insertion of a center vortex operator \eqref{eq:2dcentervortex} can be viewed as defining a partition function $\tilde Z(g^2,k)$ which is obtained from the usual partition function by modifying the couplings appearing in the pure gauge portion of the action. In particular, inserting a center vortex is equivalent to analytically continuing the couplings
\begin{equation}
g_\alpha^2 \to e^{\frac{2\pi i k}{N}n_\alpha}g_\alpha^2 
\end{equation}
in the Boltzmann weight for the plaquette $p = \star \tilde x$, where the $g_\alpha^2$ are defined as in Eq.~\eqref{eq:heatkernelweight}, while leaving them alone everywhere else. Consequently, standard clustering arguments~\cite{Munster:1981es} imply that in the hopping expansion $\tilde Z$ takes the form 
\begin{equation}
\tilde Z(g^2,k) = e^{-A \tilde F(k)} = e^{-A(F + \frac{1}{A}f(k))}\,,
\end{equation}
where $A$ is the spacetime area, which is assumed to be large,\footnote{Taking the area to be large allows us to focus on matter fluctuations on top of the trivial universe of 2d pure YM theory. Here `universe' refers to the fact that e.g. partition functions in 2d gauge theories with $\Z_N$ $1$-form symmetries decompose into contributions from $N$ disconnected sectors called universes~\cite{Hellerman:2006zs}.  For more on universes see e.g. Refs.~\cite{Pantev:2005rh,Hellerman:2006zs,Sharpe:2015mja,
Komargodski:2017dmc,Anber:2018jdf,
Anber:2018xek,Armoni:2018bga,Cherman:2019hbq,Tanizaki:2019rbk,Misumi:2019dwq,Hidaka:2019mfm,Eager:2020rra,
Komargodski:2020mxz,Cherman:2020cvw,Hidaka:2021mml,Hidaka:2021kkf, Nguyen:2021naa,Sharpe:2021srf,Cherman:2021nox}. } and in the second equality we use the fact that $\tilde Z$ differs from $Z$ by a local modification of the action, and hence should induce a sub-extensive correction to the free energy $F$. 
Here $f(k)$ is some function of $k$ mod $N$ with $f(0)=0$ and an expansion in powers of $1/N$, and $Z = e^{-A F}$, so that
\begin{equation}
\langle U_k(\tilde x)\rangle = \frac{\tilde Z(g^2,k)}{Z} = e^{-f(k)} = e^{-\left(N^2 f_2(k)+N f_1(k)+ N^0 f_0(k) + \cdots \right) } \,.
\end{equation}
On general grounds, one expects that at large $N$ gluonic degrees of freedom contribute at $\O(N^2)$ to the free energy, while matter contributions start at $\O(N)$. 
Since the pure glue sector of the theory is insensitive to the presence of the center vortex (indeed, the expectation value of the center vortex is $1$ in the pure gauge theory in infinite volume), it is natural to expect $f_2(k) = 0$ above, so that
\begin{equation}
\langle U_k(\tilde x)\rangle = e^{- N \left(f_1(k) - \frac{1}{N} f_0(k) +\cdots \right) }\,.
\label{eq:U_VEV_ansatz}
\end{equation}
The arguments above can be extended to e.g. the two-point function of $U_k$, so that it should take the form
\begin{align}
\langle U_{k}(\tilde x)^{\dag} U_k(\tilde y)\rangle = \langle U_{-k}(\tilde x)U_k(\tilde y)\rangle = e^{- N \left(c_1(k;\tilde x,\tilde y) - \frac{1}{N} c_0(k;\tilde x,\tilde y) +\cdots \right) }\,.
\label{eq:U_correlator_ansatz}
\end{align}
If quark loops were to be universally suppressed in large $N$ QCD, and $U_k$ were to remain a topological operator that generates a $\Z_N$ $1$-form symmetry, then it would have to be the case that $f_1 = f_0 = c_1 = c_0 = 0$ in the infinite volume limit, so that $\langle U_k(\tilde{x}) \rangle = 1 + \O\left(\frac{1}{N}\right)$ and $\langle U_k(\tilde{x}) U_{k}(\tilde{y}) \rangle = 1 + \O\left(\frac{1}{N}\right)$. However, as we already foreshadowed in Section~\ref{sec:quark_loop}, our calculations below  will show that $f_1 \neq 0$ and $c_1 \neq 0$.

We work in the large mass regime of the theory and expand in the hopping parameter $\kappa$. Observables depend on the combination $\kappa/m^2$, and in terms of the lattice spacing $a$, $\kappa/m^2 \sim 1/(a^2m^2)$.  This means that the hopping expansion does not access the continuum-limit region $a m \ll 1$.  
Nevertheless, the hopping expansion is a very useful tool for our purposes, because if a $\Z_N$ $1$-form symmetry were to emerge at large $N$, then it should already emerge for $a m \gg 1$ where the hopping expansion is valid.
%The large mass hopping expansion can be formulated using cluster expansion methods.
At the operational level, computations in the hopping expansion reduce to relatively simple computations in pure YM theory, which we review in Appendix~\ref{sec:2dYM}. When we take the large $N$ limit, we perform the hopping expansion before taking $N\to \infty$, with both the 't Hooft coupling $\lambda = g^2N$ and $\kappa$ held fixed. Possible large $N$ phase transitions separating the large and small mass regimes are not relevant for our analysis, which will be performed in the large mass regime. 

We will use this model for several purposes.  First, we will verify that quark loop contributions to typical gluonic observables are suppressed in the large $N$ limit, despite the fact that the glue sector of the continuum action is essentially trivial (there are no propagating gluons in two dimensions). Second, we will show that quark loop contributions to correlation functions that involve the putative generators of the $1$-form symmetry are \emph{not} suppressed in general.  This yields a very concrete setting in which we can see that there is no emergent $1$-form symmetry at large $N$.  

%%%%%%%%%%%%%%%
\subsection{Suppression of quark loops at large $N$}
%%%%%%%%%%%%%%%%%
In the following we let $\langle \cdot \rangle_0$ denote expectation values in pure 2d YM theory. In the large volume limit the fundamental Wilson loop $W_{\F}(C) = \tr \mathcal P \prod_{\ell \in C} u_\ell$ has an area law in pure YM (see Appendix~\ref{sec:2dYM} for details):
\begin{equation} \label{eq:2dWilson}
\frac{1}{N}\langle  W_{\F}(C)\rangle_0 = e^{-\sigma A}\,, \quad \sigma = g^2 c_{\F} = g^2\frac{N^2-1}{2N}\, .
\end{equation}
To get a feeling for the effects of dynamical matter on this basic observable, let us compute the first non-trivial corrections to \eqref{eq:2dWilson} in the hopping expansion, arising at $\O(\kappa^4)$. At this order hopping loops are single plaquettes, summed over all positions on the lattice whose area $A$ we will take to infinity. The evaluation of $\langle \frac{1}{N} W_{\F}(C) \rangle$ requires calculating the numerator with the operator insertion, which we denote $\ldangle \cdot \rdangle$, and the denominator $Z$, $\langle \cdot \rangle = Z^{-1}\ldangle \cdot \rdangle$. The partition function expanded to $\O(\kappa^4)$ is
\begin{align}\label{eq:Zhopping}
Z &=  Z_0 + \left(\frac{\kappa}{m^2}\right)^4 A \ldangle \tr(u_p + u_p^\dagger) \rdangle_0 +\O(\kappa^6) =   1 + \left(\frac{\kappa}{m^2}\right)^4 A \, 2N\, e^{-g^2c_{\F}}+\O(\kappa^6)\,,
\end{align}
where $A$ is the area of spacetime. To get the second equality we used the fact that the pure 2d gauge theory partition function on a surface of genus $\mathfrak g$ is $Z_0 = \sum_{\alpha} d_\alpha^{2-2\mathfrak g}e^{-g^2 c_{\alpha} A}$, and at large $A$ (with $A \to \infty$ before $N\to \infty$) this sum is dominated by the trivial representation, so that $Z_0 = 1$. 

Contributions to the numerator of the expectation value can be separated into hopping loops which are either outside or inside the Wilson loop:
\begin{align} 
\!\!\ldangle \frac{1}{N}  W_\F(C)\rdangle =  &\ldangle \frac{1}{N} W_\F(C) \rdangle_0 + \left(\frac{\kappa}{m^2}\right)^4 \sum_p \ldangle \frac{1}{N} W_\F(C) \tr(u_p + u_p^\dagger)\rdangle_0  \label{eq:WilsonNumeratorHopping} \\
=  e^{-g^2 c_{\F} A[C]} &+ \left(\frac{\kappa}{m^2}\right)^4 (A-A[C])\, e^{-g^2 c_{\F} A[C]} 2N\, e^{-g^2c_{\F}} \nonumber
\\
&+ \left(\frac{\kappa}{m^2}\right)^4 A[C]\, e^{-g^2c_{\F} (A[C]-1)} \sum_\rho (N_{\F \F}^\rho +  N_{\F\overline{\F}}^\rho) \frac{d_\rho}{N} \, e^{-g^2c_\rho} +\O(\kappa^6) \nonumber \\
= \ldangle \frac{1}{N} W_\F(C) \rdangle_0  \Bigg[ 1 &+ \left(\frac{\kappa}{m^2}\right)^4 (A-A[C])\,  2N\, e^{-g^2c_{\F}} \nonumber
\\
&+ \left(\frac{\kappa}{m^2}\right)^4 A[C] \sum_\rho (N_{\F \F}^\rho +  N_{\F\overline{\F}}^\rho) \frac{d_\rho}{N} \, e^{-g^2 (c_\rho-c_{\F})} +\O(\kappa^6)\Bigg]\,, \nonumber
\end{align}
where $A[C]$ is the area of $C$. 
In the normalized Wilson loop expectation value, part of the `outside' contributions cancel against the corresponding terms in the denominator of the expectation value \eqref{eq:Zhopping}, and the result is 
\begin{align} \label{eq:expandedW} 
\frac{1}{N} \langle W_\F(C)\rangle 
= \frac{1}{N}\langle W_\F(C)\rangle_0\Bigg\{ 1 + A[C]\left(\frac{\kappa}{m^2}\right)^4  \Bigg[\sum_\rho (N_{\F \F}^\rho &+  N_{\F\overline{\F}}^\rho)\frac{d_\rho}{N}\, e^{-g^2 (c_\rho-c_{\F})} \\
&- 2N\, e^{-g^2c_{\F}} \Bigg] +\O(\kappa^6)\Bigg\} \,. \nonumber
\end{align}

We can now examine the large $N$ behavior of the above expression. The dimensions and Casimirs of the representations appearing in the above sum are given in Table~\ref{table:rep_data}, and one can see that a generic term in Eq.~\eqref{eq:expandedW} is $\O(N)$. However, keeping only the leading terms in the Casimirs, we see that the $\O(N)$ and $\O(1)$ parts of the hopping contributions cancel, since
\begin{equation}
\frac{d_{\text{S}} + d_{\text{AS}} + d_{\text{adj}}}{N} - 2N = \frac{N^2+N}{2N} + \frac{N^2-N}{2N} + \frac{N^2-1}{N} - 2N = -\frac{1}{N},
\end{equation}
and leaves
\begin{align} \label{eq:expandedWlargeN}
\frac{1}{N} \langle W_\F(C)\rangle = \frac{1}{N}\langle W_\F(C)\rangle_0\Bigg\{1 + A[C] \left(\frac{\kappa}{m^2}\right)^4  \frac{2}{N}\sinh\left(\frac{\lambda}{2}\right) +\O(\kappa^6)\Bigg\}\,.
\end{align}
Given the overall factor of $A[C]$ in front of the leading correction term, it is natural to expect that summing over multiple insertions of single-plaquette hopping loops will lead to exponentiation of the above result. Hence, Eq.~\eqref{eq:expandedWlargeN} encodes a $\O(1/N)$ correction to the confining string tension in Eq.~\eqref{eq:2dWilson}. This explicit calculation is consistent with the general lore that the contribution of quark loops (hopping loops in the present formalism) to gluonic observables are $1/N$ suppressed.

%%%%%%%%%%%%%%%%%%%%%%%
\begin{table}[t]
\begin{center}
  \begin{tblr}{ |Q[c,m] || Q[c,m] | Q[c,m] | Q[c,m] | Q[c,m] | Q[c,m] |}
    \hline
    $R$ & $\mathbbm{1}$ & $\textrm{F}$ & $\textrm{S}$ & $\textrm{AS}$ & $\textrm{adj}$ \\ 
    \hline
    $d_R$ & $1$ & $N$ & $\frac{N^2 + N}{2}$ & $\frac{N^2 - N}{2}$ &$N^2-1$ \\ \hline 
    $c_R$ & $0$ & $\frac{N^2-1}{2N}$ & $ \frac{N^2+ N-2}{N}$ & $ \frac{N^2- N-2}{N}$ & $N$ \\
    \hline
  \end{tblr}
  \label{table:rep_data}
  \caption{Dimensions and quadratic Casimirs for the fundamental $\textrm{F}$, two-index symmetric $\textrm{S}$ and anti-symmetric $\textrm{AS}$, and adjoint $\textrm{adj}$ $SU(N)$ representations. }
\end{center}
\end{table}
%%%%%%%%%%%%%%%%%%

Of course, dynamical matter is also responsible for the qualitatively new effect of screening. The leading contribution to the perimeter law in the theory with matter comes at $\O(\kappa^{L})$, where $L$ is the perimeter of $C$. At this order the hopping loop can completely screen the Wilson loop, giving the large $N$ result
\begin{align}
\lim_{|C|\to\infty} \frac{1}{N}\langle W_{\F}(C) \rangle = \frac{1}{N} e^{-\mu L},
 \label{eq:2dQCD_wilson}
\end{align}   
where $\mu = \log(m^2/\kappa)$. This clearly exhibits the expected large $N$ behavior. The $1/N$ suppression of the perimeter-law term relative to the area law term~\eqref{eq:2dWilson} is again consistent with the fact that quark loop contributions are $1/N$ suppressed within this paradigmatic gluonic observable. 

Similarly, the leading contribution to the Polyakov loop expectation value, which is strictly zero in the pure gauge theory, comes from summing over the positions of a single oppositely-oriented non-contractible hopping loop. 
Let us first calculate the result on $S^1_L \times S^1_\beta$, and then take the $L\to\infty$ limit. The leading order contribution to the expectation value of the Polyakov loop wrapping the $\beta$ cycle in QCD can be related to the zero-momentum correlator of Polyakov loops in pure YM, 
\begin{equation}  
\frac{1}{N}\langle \tr P \rangle = \frac{1}{N}\left(\frac{\kappa}{m^2}\right)^\beta \sum_{y}\langle \tr P^\dagger(y) \tr P(0)\rangle_0\,.
\end{equation}
The correlator in question is easily calculated using the heat-kernel action, 
\begin{align}
\langle \tr P^\dagger(y) \tr P(0)\rangle_0 = \frac{1}{Z_0} \sum_{\alpha,\gamma} N_{\alpha \overline{\F}}^\gamma N_{\gamma \F}^\alpha \, e^{-g^2c_\alpha \beta y} e^{-g^2c_\gamma \beta (L-y)}\,,
\end{align}
where $Z_0$ is the pure gauge theory partition function and $N_{\alpha\overline{\F}}^\gamma$ denotes the multiplicity of $\gamma$ in $\alpha\otimes\overline{\F}$. Summing over $y \in \{1,\ldots,L\}$, we find
\begin{align}
\frac{1}{N}\langle \tr P \rangle = \frac{1}{N}\left(\frac{\kappa}{m^2}\right)^\beta\frac{1}{Z_0} \sum_{\alpha,\gamma} (N_{\alpha \overline{\F}}^\gamma)^2\, \frac{e^{-g^2c_\alpha\beta L}-e^{-g^2c_\gamma\beta L}}{1-e^{-g^2(c_\gamma-c_\alpha)\beta}} \,.
\end{align}
Taking the limit $L\to\infty$, $Z_0 \to 1$, we only keep the contributions from $\alpha = \mathbbm{1}, \gamma = \overline{\F}$ and $\gamma = \mathbbm{1}, \alpha = \F$. We then arrive at the Polyakov loop expectation value on $\RR\times S^1_\beta$,
\begin{align} \label{eq:2dpolyakov} 
\frac{1}{N}\langle \tr P \rangle  = \frac{1}{N}e^{-\mu\beta}\, \text{coth}\left(\frac{\lambda\beta}{4}\right)\,. 
\end{align}
As a result, the properly normalized expectation value $\tfrac{1}{N}\langle \tr P\rangle$ vanishes at large $N$. 
Finally, we can verify that properly normalized open Wilson lines have $\mathcal{O}(1)$ expectation values. 
If we take a straight open Wilson line between two sites separated by a distance $|x|$ on the lattice, the leading contribution is simply 
\begin{equation} \label{eq:openline}
\frac{1}{N} \langle \phi^\dagger(x) W_{x,0}\,  \phi(0) \rangle =\frac{1}{m^2} \left(\frac{\kappa}{m^2}\right)^{|x|} =\frac{1}{m^2} e^{-\mu |x|}. 
\end{equation}

The center-symmetric expectation values \eqref{eq:2dQCD_wilson}, \eqref{eq:2dpolyakov} are crying out for an explanation in terms of a symmetry. However, the existence of non-trivial open Wilson lines means that the codimension-2 center-vortex operators defined in Eq.~\eqref{eq:2dcentervortex} cannot be topological at large $N$.

%%%%%%%%%%%%%%%
\subsection{Non-suppression of quark loops at large $N$}
\label{sec:nonsuppression}
%%%%%%%%%%%%%%%%%  

We now examine the effects of fundamental matter fields on correlation functions of the center-vortex operators defined in Eq.~\eqref{eq:2dcentervortex}. 
As in the case of the Wilson loop, one might expect the contributions of matter loops to be negligible at large $N$, at the very least in sectors of the theory with vanishing overlap with open Wilson lines. 
For instance, one might expect the correlation functions of center-vortex operators and closed Wilson loops to reduce to their pure gauge theory values. 
We begin by being less ambitious, and simply examine whether the expectation values of the center-vortex operators remain equal to $1$ (their pure gauge theory value) in the large $N$ limit. 
We will see that this turns out not to be the case, indicating that the standard large $N$ counting fails for the would-be generators of the $\Z_N$ $1$-form symmetry.  This result fits with the general argument given in Section~\ref{sec:quark_loop}.

As above, we start with the numerator of the expectation value of the center-vortex operator evaluated to the leading non-trivial order in the hopping expansion. The contributions of single-plaquette hopping loops depend on whether they encircle the center-vortex or not, 
\begin{align}
\ldangle U_k(\tilde x) \rdangle &= \ldangle U_k(\tilde x) \rdangle_0 + \left(\frac{\kappa}{m^2}\right)^4\sum_p \ldangle U_k(\tilde x)\tr(u_p + u_p^\dagger) \rdangle_0 +\O(\kappa^6)  \\
&= 1 + \left(\frac{\kappa}{m^2}\right)^4\,(A-1)\,2N\, e^{-g^2c_{\F}} + \left(\frac{\kappa}{m^2}\right)^4\,N(\omega^k + \omega^{-k})\, e^{-g^2 c_{\F}} +\O(\kappa^6) \,.\nonumber
\end{align}
Dividing by the partition function \eqref{eq:Zhopping}, we find
\begin{align}
\langle U_k(\tilde x) \rangle &= 1 - \left(\frac{\kappa}{m^2}\right)^4 \, 2N\,e^{-g^2c_{\F}} \left(1-\cos\left(\frac{2\pi k}{N}\right)\right) \,.
\label{eq:Uk_VEV_leading}
\end{align} 
Due to the overall factor of $N$ in the $\O(\kappa^4)$ term, this leading correction is generically $\O(N)$.  Eq.~\eqref{eq:Uk_VEV_leading} is enough to see that that the naive large $N$ counting, which states that quark loops are $1/N$-suppressed for gluonic observables, fails for the center-vortex operators.  Moreover, as discussed around Eq.~\eqref{eq:VEV_constraint} above, the expectation values of $\ZZ_N$ symmetry operators are constrained to be $N$th roots of unity. We can then conclude, already from the leading order result in the hopping expansion Eq.~\eqref{eq:Uk_VEV_leading}, that the center vortex operators $U_k(\tilde x)$ cannot generate a $\ZZ_N$ 1-form symmetry at large $N$. Moreover, examination of the next orders shows that the value of the $\O(\kappa^8 N^2)$ term is consistent with the general exponentiation ansatz from \eqref{eq:U_VEV_ansatz}:
\begin{align}
\langle U_k(\tilde x) \rangle &=  \exp\left(- N f_1(k)\right)\,.
\label{eq:Uk_VEV_exponentiated}
\end{align} 
with
\begin{align}
    f_1(k) = \left(\frac{\kappa}{m^2}\right)^4 \, 2\,e^{-g^2c_{\F}} \left(1-\cos\left(\frac{2\pi k}{N}\right)\right) + \mathcal{O}(\kappa^6)\,.
\end{align}
We have thus learned that the expectation values of center vortex operators with $k \sim \O(N)$ vanish \emph{nonperturbatively} in $1/N$ as $N\to \infty$.  At the same time, the $U_k(\tilde{x})$ operators with $k \sim \mathcal{O}(1)$, which become effectively trivial operators at large $N$, have unit expectation values up to $1/N$ corrections, as one would expect from our discussion in the preceding sections.  Finally, center vortex operators with $k \sim \O(\sqrt{N})$ have $\mathcal{O}(1)$ expectation values which deviate from $1$ for generic small values of $\kappa$.  These results are clearly inconsistent with the appearance of a  $\mathbb{Z}_N$ $1$-form symmetry at large $N$.

It is tempting to try to repair this issue by defining rescaled operators 
\begin{equation}
V_k(\tilde x) = \frac{U_k(\tilde x)}{|\langle U_k(\tilde x)\rangle|}     
\end{equation}
which are constructed to have unit expectation values at all values of $N$.  
The operators $V_k$ are well-defined at finite $N$, but it is not obvious that their correlation functions have smooth large $N$ limits given the somewhat singular ``$0/0$'' behavior of their definition as $N \to \infty$. This motivates exploring the properties of the two-point correlation function of $V_k$.  Suppose $\tilde x$ and $\tilde y$ are sites on the dual lattice connected by a straight path of length $r$.  Then the discussion leading to \eqref{eq:U_correlator_ansatz} implies that
\begin{align}
\langle V_k^\dagger(\tilde x) V_k(\tilde y) \rangle = e^{-N\left[c_1(k; r) - 2f_1(k)\right] - N^0 c_{0}(k; r) + \cdots}
\label{eq:Utilde_correlator_ansatz}
\end{align}
for large $N$. Let us make the assumption that the vacuum state of 2d large $N$ QCD enjoys cluster decomposition for all correlation functions of local operators.   Then cluster decomposition implies that $\lim_{r\to\infty} c_1(k;r) = 2f_1(k)$, $\lim_{r\to\infty} c_0(k;r) = 0$. However, if $c_1(k;r)$ deviates from $2f_1(k)$ as a function of $r$, then at large $N$ $\langle V_k^\dagger(\tilde x) V_k(\tilde y) \rangle$ will not have a smooth large $N$ limit.  If $c_1(k;r)$ depends on $r$, then at large $N$ we would get a function which diverges wherever $c_1(k;r) < 2f_1(k)$, is zero wherever $c_1(k;r) > 2f_1(k)$, and becomes finite when $c_1(k;r) = 2f_1(k)$.  This bizarre behavior would be (a) inconsistent with any attempt to interpret $\tilde U_k$ as topological operators and (b) signal a failure of cluster decomposition at large $N$, since $\langle V_k(\tilde{x})^{\dag} V_k(\tilde{y})\rangle$ would not factorize for any finite $r = |\tilde{x}-\tilde{y}|$, except at particular values $r_0$ which satisfy $c_1(k;r_0) = 2f_1(k)$ and $c_0(k;r_0) = 0$.  So to get a smooth large $N$ limit for $V_k$ we would need to find that $c_1(k;r) = 2f_1(k)$ for \emph{all} $r$.  Moreover, to interpret $V_k$ as topological operators at large $N$, we would further need $c_0(k;r) = 0$.

We can test this scenario using the hopping expansion.  First consider the un-normalized operators $U_k$. The leading diagram in the hopping expansion contributing to the two-point connected correlator $\langle\cdot \cdot \rangle_c = \langle\cdot \cdot \rangle - \langle\cdot\rangle \langle \cdot \rangle$ occurs at $\O(\kappa^{2r+4})$, and consists of a single rectangular hopping loop $W_{\F}(C)$ and its conjugate. All other diagrams to this  order in the hopping expansion only contribute to the disconnected part of the correlator. The contribution to the two-point function from these individual rectangular hopping loops is
\begin{align}
\label{eq:connected_loop}
\left(\frac{\kappa}{m^2}\right)^{2r+4}\Bigg\{& \langle U_k^\dagger(\tilde x)U_k(\tilde y) (W_{\F}(C) + W_{\F}^\dagger(C))\rangle_0\\
&+ \left(2(r+1)-1\right) \left[\langle U_k^\dagger(\tilde x)(W_{\F}(C)+W_{\F}^\dagger(C))\rangle_0 \langle U_k(\tilde y) \rangle_0 + \text{h.c.}\right] \nonumber \\
& - (4(r+1)-1)\langle U_k^\dagger(\tilde x) U_k(\tilde y) \rangle_0\langle W_{\F}(C)+W_{\F}^\dagger(C) \rangle_0 \Bigg\}\,, \nonumber
\end{align}
where the contour $C$ encloses the points in each given pure gauge theory expectation value. The contribution in \eqref{eq:connected_loop} can be used to deduce the leading term in the hopping expansion of the two-point function of the normalized operators $V_k$:
\begin{align} \label{eq:centervortexcorrelator}
\langle V_k^\dagger(\tilde x) V_k(\tilde y)\rangle &=1+\left(\frac{\kappa}{m^2}\right)^{2r+4}4N\, e^{-g^2c_\F (r+1)}\left(1-\cos\left(\frac{2\pi k}{N}\right)\right)\, +\mathcal{O}\left(\frac{\kappa}{m^2}\right)^{2r+6}\,.
\end{align}
This is consistent with an expansion of Eq.~\eqref{eq:Utilde_correlator_ansatz} with a $c_1(k;r)$ 
which \emph{does} depend on $r$.  At large but finite $N$, and small $\kappa$, the correlator is exponentially large in $N$ for $r \sim N^0$, and cluster decomposition sets in only for $r \sim \log N$.  If we take the 't Hooft large $N$ limit with fixed $r$, then the operators $V_k$ do not obey cluster decomposition.  This means that $V_k(\tilde{x})$ with $k\sim \O(N)$ have singular correlation functions at large $N$.  Our results also imply that $V_k(\tilde{x})$ with $k\sim \O(\sqrt{N})$ have non-singular and also non-topological correlation functions at large $N$.  We thus conclude that $V_k(\tilde{x})$ cannot be interpreted as topological operators that generate a $\Z_N$ $1$-form symmetry in large $N$ QCD. 

%%%%%%%%%%%%%%%%
\section{Symmetry generators and large $N$ as a classical limit}
\label{sec:classical}
%%%%%%%%%%%%%%%%
 
In the previous sections we gave arguments against the existence of codimension-2 operators in QCD which act on Wilson loops and are topological in the large $N$ limit. In particular, we encountered the fact that the $\ZZ_N$ 1-form symmetry generators $U_k$ of pure YM theory with $k \sim N$ do not have smooth large $N$ behavior when fundamental quarks are present in the theory. This may be surprising given that such operators are purely gluonic. Here we consider our observation from the perspective of Ref.~\cite{Yaffe:1981vf}, which formulates the large $N$ limit in terms of the classical dynamics of sufficiently well-behaved observables.\footnote{We thank L.~Yaffe for illuminating discussions on this approach.}

Reference~\cite{Yaffe:1981vf} introduced a general method for recasting large $N$ limits of quantum theories as certain classical limits. This formalism can be used to study a class of `classical operators' satisfying certain properties.  Crucially, the correlation functions of such operators must factorize in the large $N$ limit. This immediately implies that operators which transform states in the classical phase space, such as symmetry generators which multiply charged operators by $\mathcal{O}(1)$ linking-dependent phases, are not `classical operators' in the sense defined in Ref.~\cite{Yaffe:1981vf}. One should not expect the correlation functions of such operators to satisfy the standard $N$ counting rules.

These considerations already apply to the generators of an exact symmetry, making it natural to expect that these same operators will not be well-behaved at large $N$ if the symmetry is explicitly broken. While this helps contextualize our results above, it also does not provide a satisfying alternative to thinking about exact, let alone emergent, symmetries in large $N$ gauge theories. In this paper we have approached the problem using the language of generalized global symmetries, where all symmetries are understood in terms of topological operators.  That being said, it is possible that some formalization of the notion of symmetry based on the classical description of large $N$ quantum theories along the lines of Ref.~\cite{Yaffe:1981vf} may give further insights into the apparent discrepancy between symmetries and selection rules highlighted in this paper.  
  
%%%%%%%%%%%%%%%%
\section{Conclusions}
\label{sec:conclusions}
%%%%%%%%%%%%%%%%

$SU(N)$ pure YM theory obeys many Wilson loop selection rules. It also has a set of $N$ codimension-$2$ topological operators with $\Z_N$ fusion rules and a $\Z_N$ action on Wilson loops. The existence of these operators means that $SU(N)$ pure YM theory has a $\Z_N$ $1$-form symmetry, which can be used to explain selection rules for Wilson loops charged under the symmetry. Large $N$ QCD obeys the same selection rules, which makes it tempting to believe that it has the same $1$-form symmetry as pure YM.   However, we have found that large $N$ QCD does \emph{not} have a $1$-form symmetry, because the necessary codimension-$2$ topological operators do not exist.

We gave three arguments for this conclusion.  First, in Section~\ref{sec:endability} we observed that the existence of open fundamental-representation Wilson line operators in large $N$ QCD yields a very general argument that it cannot have codimension-$2$ topological operators with a $\Z_N$ action on Wilson loops.  Next, in Section~\ref{sec:quark_loop}, we observed that quark loop contributions to correlation functions of the would-be topological operators are not suppressed in large $N$ QCD, which indeed is necessary for the arguments in Section~\ref{sec:endability} to be consistent.  The basic point is that among the set of $N$ would-be topological operators $U_k$, the ones that remain non-trivial at large $N$ have $k \sim N$, and the standard quark-loop suppression argument do not apply to such operators.  These general arguments about the absence of a $\Z_N$ $1$-form symmetry in large $N$ QCD apply both on the lattice and in the continuum limit, and in any dimension $d \ge 2$ where QCD can be defined. In Section~\ref{sec:classical} we note that this observed quark-loop non-suppression is not unexpected if one views the large $N$ limit as a kind of classical limit. Third, in Section~\ref{sec:2dQCD} we verified these arguments by direct analytic calculations in the simplest tractable example, namely 2d large $N$ scalar QCD on a lattice, treated using a hopping expansion.  Here we took the generators of the $\Z_N$ $1$-form symmetry in pure gauge theory as the would-be codimension-$2$ topological operators in large $N$ QCD, and examined their expectation values and correlation functions.  These operators do not become topological at large $N$.   The properties of these non-topological operators in 2d large $N$ QCD are somewhat peculiar, and may be of independent interest to some readers.  In future work it may be interesting to examine the large $N$ behavior of thin center-vortex operators in other spacetime dimensions, as well as near the continuum limit. 

Our results are surprising because it is generally expected that selection rules are the consequence of symmetries, while here we have shown that large $N$ QCD has some selection rules that are not the consequence of any known symmetry. Large $N$ QCD does have a $\Z_N$ center symmetry, see Section~\ref{sec:center_sym}, and this symmetry can be used to explain some (but not all!) selection rules.  A curious by-product of this set of observations is that center symmetry cannot always be thought of as a special case of a $1$-form symmetry.  Center symmetry is indeed a special case of a $1$-form symmetry in pure YM theory, as we discuss in Section~\ref{sec:center_vs_one_form}, but not in large $N$ QCD.

We should also note that in the large $N$ limit, YM theory obeys many Wilson loop selection rules which cannot be directly explained by its $\Z_N$ $1$-form symmetry either.  For example, in the large $N$ limit the adjoint Polyakov loop expectation value vanishes:
\begin{align}
   \left\langle \frac{1}{d_{\rm adj}} \tr_{\rm adj} P(\vec{x}) \right\rangle = 0 \,.
   \label{eq:adj_vanishing}
\end{align}
This is not a direct consequence of the $\Z_N$ $1$-form symmetry because the adjoint Polyakov loop is neutral under this symmetry.  Nevertheless, the selection rule holds because one can apply large $N$ factorization to the adjoint trace, and then use the  $\Z_N$ $1$-form symmetry on the factorized pieces, which involve fundamental-representation traces.  Since large $N$ Wilson loop expectation values in YM coincide with those in large $N$ QCD, this selection rule also holds in large $N$ QCD.   This selection rule (and an infinite number of related ones) also poses a challenge to the idea that selection rules should be explained by symmetries.  
   
An interesting recent observation by Nguyen, Tanizaki, and \"Unsal is that in $d=2$ pure $SU(N)$ YM theory the selection rule above can be explained by an exact \emph{non-invertible} $1$-form symmetry~\cite{Nguyen:2021naa}.  
The existence of this exotic symmetry can be used to explain why \eqref{eq:adj_vanishing} holds even at \emph{finite} $N$ in 2d pure YM on $\R^2$.\footnote{Invertible symmetries lead to finite-volume selection rules, while non-invertible symmetries generically only lead to selection rules in infinite-volume limits; see Appendix~\ref{sec:selectionrules}.}  
This makes it tempting to speculate that when $d>2$, pure YM has an approximate non-invertible $1$-form symmetry,\footnote{There have been many recent discussions of non-invertible symmetries in $d>2$, see e.g.~\cite{Rudelius:2020orz,Nguyen:2021yld,Koide:2021zxj,Heidenreich:2021xpr,Choi:2022jqy,Kaidi:2021xfk,Choi:2022zal,Cordova:2022ieu,Antinucci:2022eat}.} in the sense of having non-invertible codimension-$2$ operators which are topological up to $1/N$ corrections, with some appropriate action on Wilson loops.  Unfortunately this scenario can now be ruled out by a repetition of any of the  arguments against the large $N$ emergence of codimension-$2$ topological operators in this paper. For example, when $d=2$ pure YM expectation values of open adjoint Wilson lines vanish identically, but when $d >2$ these expectation values do not vanish in YM theory, nor do they vanish in QCD in $d \ge 2$.  The Rudelius-Shao argument on endability then implies that codimension-$2$ topological operators that would explain Eq.~\eqref{eq:adj_vanishing} cannot exist.
   
We see two possible general interpretations of our results. While it would be rather startling, perhaps quantum field theories can enjoy robust selection rules that are not the consequence of any symmetry principles.  If so, it would be nice to develop a systematic understanding of when this can occur.  So far the only (known) examples of selection rules which are not (apparently) explained by symmetries occur in large $N$ gauge theories.    Alternatively, perhaps the large $N$ selection rules discussed above can be explained by some generalization of generalized global symmetries.  However, while the original generalized symmetry framework of Ref.~\cite{Gaiotto:2014kfa} has been generalized\footnote{For more extensive references, please see the review articles  Refs.~\cite{Cordova:2022ruw,McGreevy:2022oyu}.} in various directions, such as to higher groups, see e.g.~\cite{Benini:2018reh,Cordova:2018cvg}, non-invertible topological operators, see e.g.~\cite{Bhardwaj:2017xup,Chang:2018iay,Rudelius:2020orz} and to condensation defects, see e.g.~\cite{Roumpedakis:2022aik,Choi:2022zal}, for now there is no known generalization of the type necessary to explain the large $N$ selection rules discussed here.

\acknowledgments
We are very grateful to Mendel Nguyen, Yuya Tanizaki, and Mithat \"Unsal for intensive discussions at the beginning of this project, and thank Yuya and Mithat for comments on a draft of this paper.  We are also indebted to Gerald Dunne, Zohar Komargodski, Shu-Heng Shao, Sahand Seifnashri, Matthew Strassler, and Larry Yaffe for very helpful remarks. A.C. and T.J. thank the Simons Center for Geometry and Physics for its hospitality during the ``Confronting Large N, Holography, Integrability and Stringy Models with the Real World" program in Spring 2022, where some of this work was done, and A. C. is grateful to the members of the Simons Collaboration on Confinement and QCD Strings for helpful questions during a seminar. M.N. was partly supported by the National Science Foundation Graduate Research Fellowship under Grant No. 1842400. This work was also supported in part by the University of Minnesota. 

\appendix

%%%%%%%%%%%%%%%%%
\section{Selection rules from 1-form symmetries}
%%%%%%%%%%%%%%%%%
\label{sec:selectionrules}
In this section we review how $1$-form symmetries lead to selection rules for correlation functions of Wilson loops in a QFT in $d$ spacetime dimensions.  While some expert readers may find our discussion in this Appendix somewhat slow, we could not find a pedagogical presentation of this material in the literature, and hope that other readers may find it helpful.  Almost of all of our remarks below are not new, but it is possible that the distinctions between selection rules associated with invertible and non-invertible symmetries that we highlight toward the end of this Appendix may not yet be widely appreciated.

Suppose that spacetime is a closed manifold $M_d$, and we have a QFT with a group-like one-form symmetry  $G^{(1)}$.  One-form symmetries have to be abelian and they are generically discrete, so we can take $G^{(1)} = \Z_N$ without loss of generality.   This means that there exist $N$ operators $U_{k}(M_{d-2})$ with the properties 
\begin{align}
    U_{k}(M_{d-2})U_{k'}(M_{d-2}) = U_{k + k' \textrm{ mod } N}(M_{d-2})
\end{align}
and 
\begin{align}
    \langle U_{k}(M_{d-2}) W(C) \rangle = 
    e^{2\pi i k/N} \langle U_{k}(\tilde{M}_{d-2}) W(C)\rangle
     \label{eq:1sym_action}
\end{align}
where $M_{d-2}$ links the closed simple curve $C$ while $\tilde{M}_{d-2}$ does not, and we have assumed that the charged line operator $W(C)$ has unit charge under the $1$-form symmetry.   We also assume that $M_{d-2}$ is contractible to a point, so that
\begin{align}
    \langle U_{k}(M_{d-2})\rangle = 1
\end{align}
when $d>2$. In $d=2$, where $M_{d-2}$ is a point $x$ in the first place, inserting $W(C)$ necessarily splits spacetime into two regions (which are called `universes' in this context).  This means that expectation values of $U_k(x)$ can depend on how we insert the charged operators $W(C)$ at infinity.  Consistency with the $\Z_N$ fusion rules implies that
\begin{align}
    \langle U_{k}(M_{d-2})\rangle = e^{i 2\pi k \ell/N}
\end{align}
for some integer $\ell$ (defined mod $N$).

\begin{figure}[t]
\begin{center}
\includegraphics[width=0.8\textwidth]{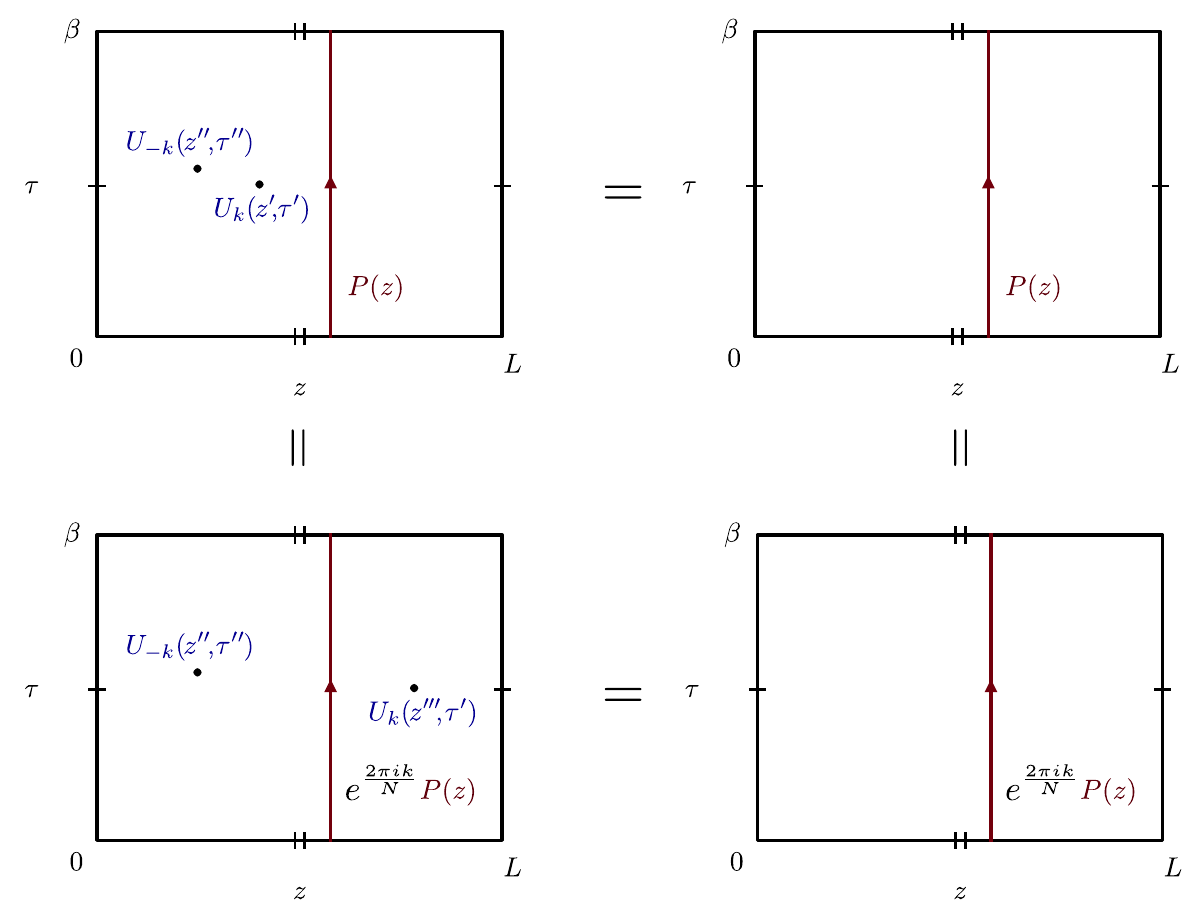}
\end{center}
\caption{To prove a finite-volume selection rule for a non-contractible line operator $P$ charged under an invertible $1$-form symmetry, one can consider a correlation function of $P$ and a generator $U_k$ of the $1$-form symmetry with its inverse $U_{-k}$.  In the figure we assume the spacetime manifold is $M_{d-2} \times S^1_L \times S^1_{\beta}$, $P$ wraps around $S^1_{\beta}$ localized at $\vec{x} \in M_{d-2}$ and a point $z \in S^1_L$, while the symmetry generators wrap around $M_{d-2}$ and are point-like on $S^1_L \times S^1_{\beta}$.  We have suppressed the dependence on $M_{d-2}$ in the figure to reduce clutter.}
\label{fig:polyakov_loop_selection}
\end{figure}

Let us call $W(C)$ Wilson loops for simplicity.  These line operators could be contractible on $M_d$, or they could be non-contractible.  In the latter case it is traditional to call them Polyakov loops.  The existence of topological operators with the properties above implies a selection rule for Polyakov loops.  To see this, suppose that the spacetime is  $M_d = M_{d-2} \times S^1_{L}\times S^1_{\beta}$ and denote points on $S^1_L$ and $S^1_{\beta}$ by $z$ and $\tau$ respectively.  We will consider a Polyakov loop $P(\vec{x},z) = W(S^1_{\beta}; \vec{x},z)$, where $\vec{x} \in M_{d-2}$.  Now consider
 \begin{align}
     \langle U_{-k}(M_{d-2}; z'',\tau'') U_{k}(M_{d-2}; z',\tau')  P(\vec{x},z) \rangle  \,,
     \label{eq:polyakov_selection_correlator}
 \end{align}
where the charge operators wrap $M_{d-2}$ and are point-like on $S^1_L \times S^1_{\beta}$. First, let us suppose that $z'',\tau''$ is close to $z', \tau'$, as  illustrated in the top left panel of Fig.~\ref{fig:polyakov_loop_selection}.  Then we can use the topological property of the $U_k$ operator to fuse it into $U_{-k}$ from the `right' and get the unit operator, so that 
 \begin{align}
     \langle U_{-k}(M_{d-2}; z'',\tau'')  U_{k}(M_{d-2}; z',\tau') P(\vec{x},z) \rangle = \langle P(\vec{x},z) \rangle  \,.
 \end{align}
 On the other hand, we can use the topological property of $U_k$ to move it past $P(\vec{x},z)$, at the cost of multiplying Eq.~\eqref{eq:polyakov_selection_correlator} by $e^{2\pi i k/N}$, and then use periodicity in $S^1_L$ to merge $U_k$ into $U_{-k}$ from the `left,' giving 
  \begin{align}
    &\langle U_{-k}(M_{d-2}; z'',\tau'')  U_{k}(M_{d-2}; z',\tau') P(\vec{x},z) \rangle \\
   = e^{2\pi i k/N}& \langle   U_{-k}(M_{d-2}; z'',\tau'') U_{k}(M_{d-2}; z''',\tau') P(\vec{x}, z) \rangle 
      \\
      = e^{2\pi i k/N} &\langle P(\vec{x},z ) \rangle  \,.
 \end{align}
 This implies the selection rule 
\begin{align}
    \langle P(\vec{x},z) \rangle = 0 \,.
    \label{eq:polyakov_vanishing_appendix}
\end{align}
We should emphasize three features of the argument leading to Eq.~\eqref{eq:polyakov_vanishing_appendix}.  First, it works in finite volume, so if a Polyakov loop is charged under an invertible $1$-form symmetry then its expectation value must vanish in finite spacetime volume. Second, the argument above does not yield a selection rule for \emph{contractible} Wilson loops.  Third, the derivation above relies on the invertibility of the $U_k$ operators.   

%%%%%%%%%%
%%%%%%%%%%%%%%

To understand the implication of $1$-form symmetries for contractible Wilson loops, we need an argument with an extra assumption, namely cluster decomposition.  We will phrase the argument in such a way that it also applies to QFTs with non-invertible $1$-form symmetries, which exist at least in $d=2,3$~\cite{Nguyen:2021yld,Nguyen:2021naa}.   As a first step, suppose we have a not-necessarily-invertible topological operator $U_{\alpha}(M_{d-2})$ which acts on Wilson loops $W_R(C)$ in the representation $R$ as 
 \begin{align}
    \langle U_{\alpha}(M_{d-2}) W_R(C) \rangle = f(\alpha,R) \langle U_{\alpha}(\tilde{M}_{d-2}) W_R(C)\rangle \,.
\end{align}
Here $C$ has unit and vanishing linking with $M_{d-2}$ and $\tilde{M}_{d-2}$, respectively, $\alpha$ is some label which is not necessarily an integer, and $f(\alpha,R)$ is not assumed to have unit modulus.  To derive a selection rule, we will need to assume that our QFT is in a state which satisfies an appropriate version of the principle of cluster decomposition:
\begin{itemize}
    \item Suppose that $\mathcal{O}(M)$ and $\mathcal{O}'(N)$ are two operators supported on the closed manifolds $M \subset M_d$ and $N \subset M_d$. Suppose that the minimum distance between points in $M$ and points in $N$ is $\ell$, and that the total volume of spacetime is $V_d$.  Then the vacuum state satisfies cluster decomposition if it obeys 
    \begin{align}
       \lim_{\ell \to \infty} \lim_{V_d \to \infty} \left[\langle \mathcal{O}(M)\mathcal{O}'(N) \rangle
        - \langle \mathcal{O}(M)\rangle 
        \langle\mathcal{O}'(N) \rangle\right] \to 0 
    \end{align}
    for all $\mathcal{O}$ and $\mathcal{O}'$ which are sufficiently well-localized so that the limits above make sense.
\end{itemize}
Note that cluster decomposition only makes sense in the infinite-volume limit where the volume $V_d$ of $M_d$ diverges.  To use this to prove a selection for a Polyakov loop in e.g. a 2d QFT on $\mathbb{R} \times S^1$, consider the correlation function
\begin{align}
    \langle U_{\alpha}(x) P_R(S^1) \rangle \,.
\end{align}
If $x$ is very far from $C$, then cluster decomposition implies
\begin{align}
       \langle U_{\alpha}(x) P_R(S^1) \rangle = \langle U_{\alpha}(x)\rangle \langle P_R(S^1) \rangle \,.
\end{align}
On the other hand, we could use the topological property of $U_{\alpha}$ to first move it past $P_R(S^1)$, which picks up a factor of $f(\alpha,R)$, and then move it very far away.  Then cluster decomposition implies
\begin{align}
       \langle U_{\alpha}(x) P_R(S^1) \rangle =f(\alpha,R) \langle U_{\alpha}(x)\rangle \langle P_R(S^1) \rangle \,.
\end{align}
Assuming that $\langle U_{\alpha}(x) \rangle \neq 0$ and $f(\alpha, R) \neq 1$, we find the infinite-volume selection rule
\begin{align}
    \langle P_R(S^1) \rangle = 0 \,.
\end{align}
If there are Polyakov loops that are only charged under a non-invertible symmetry operator $U_{\alpha}$, this argument predicts that they should have vanishing expectation values in the infinite-volume limit, but not in finite volume.  This is precisely what one finds in pure 2d $SU(N)$ YM, where Polyakov loops in (non-trivial) representations with zero $N$-ality have expectation values which vanish exponentially with the spacetime volume.  

We now consider a contractible Wilson loop $W_R(C)$.  The $1$-form symmetry better not yield any selection rules which are independent of the geometry of $C$, because when $C$ is small $W(C) \sim 1+L^4\tr F_{\mu\nu}F^{\mu\nu}(x) + \cdots $ where $L$ is a length scale associated with $C$, $x$ is a point within the region bounded by $C$, and both the unit operator and $F^2$ have non-vanishing expectation values in gauge theory.  A $1$-form symmetry implies a selection rule for contractible Wilson loops $W_R(C)$ only in the limit as $C$ becomes infinitely large, and the argument uses cluster decomposition.

%%%%%%%%%%
\begin{figure}[th]
\begin{center}
\includegraphics[width=.8\textwidth]{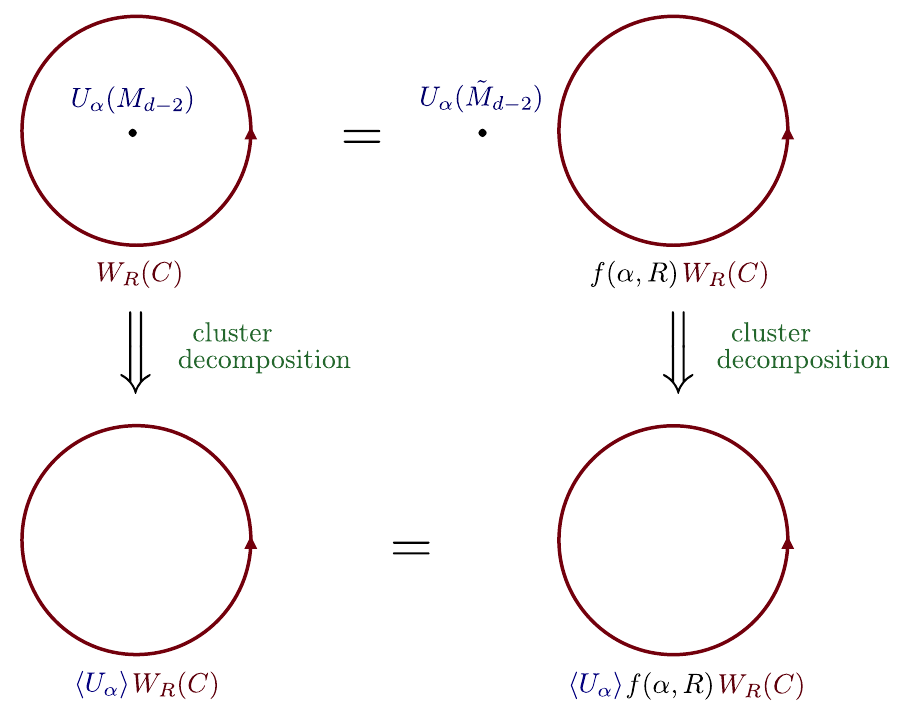}
\end{center}
\caption{To prove a selection rule for contractible closed line operators $W_R$ charged under a $1$-form symmetry, one can consider a correlation function of $W_R$ and a generator $U_{\alpha}$ of the $1$-form symmetry. Then one uses the action of the symmetry generator along with the principle of cluster decomposition.   For visual clarity, the figure assumes that  $M_{d-2}$ is point-like in the plane of the curve $C$.
To justify the invocations of cluster decomposition indicated on the right-hand side of the figure, the spacetime area must be large compared to the area bounded by $C$, so that $U_{\alpha}$ can be moved far from $C$ once it is outside of $C$. Similarly, to justify the use of cluster decomposition on the left-hand side, the area bounded by $C$ must also be large.  If $\langle U_{\alpha}\rangle \neq 0, f(\alpha,R) \neq 1$, we get the selection rule that large contractible closed line operators $W_R(C)$ must vanish in any renormalization scheme. This selection rule only holds when $C$ bounds a diverging area, in infinite spacetime volume, with an appropriate ordering of limits.  As explained in the text, the selection rule is satisfied by an area-law behavior for the contractible line operator, but not by a perimeter-law behavior. }
\label{fig:area_law_selection}
\end{figure}
%%%%%%%%%%%%%%

Consider $U_{\alpha}(M_{d-2})$ and a Wilson loop $W_R(C)$ where $C$ and $M_{d-2}$ are very far from each other and have unit linking number. Note that this requires taking a limit where $C$ itself becomes very large, see Fig.~\ref{fig:area_law_selection}. Deform $M_{d-2}$ to $\tilde{M}_{d-2}$ in such a way that the linking number of $\tilde{M}_{d-2}$ with $C$ becomes $0$, with $\tilde{M}_{d-2}$ also very far from $C$.    Then cluster decomposition implies that
\begin{align}
\langle U_{\alpha}(M_{d-2})\rangle \langle W_R(C) \rangle &= 
   f(\alpha,R)  \langle U_{\alpha}(\tilde{M}_{d-2}) \rangle \langle W_R(C)\rangle \,.
\end{align}
This yields the selection rule
\begin{align}
    \langle W_R(C) \rangle = 0
    \label{eq:contractible_selection_rule}
\end{align}
when $C$ becomes arbitrarily large.  In particular $\langle W_R(C) \rangle$ must vanish faster than a perimeter law $e^{-\mu P(C)}$ where $\mu$ is a short-distance scale and $P(C)$ is the perimeter of $C$.  This is because a perimeter-law scaling can be changed by local counterterms on $C$, which have the effect of additively shifting $\mu$.  This means that if a Wilson loop has a perimeter-law behavior, it is possible to choose a scheme where $\langle W_R(C)\rangle = 1$.  However, the selection rule in \eqref{eq:contractible_selection_rule} must hold for \emph{any} choice of counter-terms.    In YM-like theories we expect that \eqref{eq:contractible_selection_rule} is satisfied by an area-law behavior $\langle W_R(C) \rangle \sim e^{-T_R A(C)}$ where $T_R$ is the string tension and $A(C)$ is the area enclosed by $C$. 

%%%%%%%%%%%%%%%
\subsection{Comments on selection rules for non-invertible 0-form symmetries}
%%%%%%%%%%%%%%

It has recently become appreciated that non-invertible $0$-form symmetries are somewhat ubiquitous, see e.g.~\cite{Cordova:2022ruw} for a review.  They are generated by codimension-$1$ non-invertible topological defects, and our discussion above can be easily generalized to derive selection rules for non-invertible $0$-form symmetries which act homogeneously on some local operators.  The main point we want to emphasize here is that these symmetries lead to selection rules in the infinite-volume limit, because then one can take advantage of cluster decomposition and give arguments parallel to the one leading to ~\eqref{eq:contractible_selection_rule} above.  However, non-invertible $0$-form symmetries do not generically yield simple selection rules in finite volume.
\footnote{Non-invertible $0$-form symmetries can lead to finite-volume selection rules if the spacetime manifold only admits {\it contractible} symmetry generators. For example, the $1+1$d Ising model has a non-invertible $0$-form `Kramers-Wannier' symmetry line $\mathcal{N}(C)$ that leads to selection rules for the energy field $\epsilon$ and spin field $\sigma$ when the Ising model is placed on the Euclidean spacetime manifold $S^2$; see e.g. \cite{Frohlich:2004,Chang:2018iay}.  We are grateful to Shu-Heng Shao for a remark in this direction.}  Indeed, if one finds some QFT which obeys a local-operator selection rule on $\R^d$ but not on $T^{d}$, it is possible (but not necessarily true) that this selection rule is explained by a non-invertible $0$-form symmetry. 

As an illustration, consider the recent observation that 4d QED with massless fermions has a non-invertible version of chiral symmetry~\cite{Choi:2022jqy,Cordova:2022ieu}.  The operators generating this symmetry act on $\bar{\psi}_L \psi_R$ by multiplying it by rational phases.  They also have a more complicated action on 't Hooft line operators. One consequence of this non-invertible symmetry is the vanishing of the chiral condensate on $\mathbb{R}^4$. This could also be seen from the absence of instantons in QED on $\R^4$, see e.g.~\cite{Harlow:2018tng}. However, QED instantons do exist on $T^4$, and correspondingly, the chiral condensate of 4d QED on $T^4$ does not vanish.   This difference in behavior of $\bar{\psi}_L \psi_R$ on $\mathbb{R}^4$ versus $T^4$ would be inconsistent with an invertible $U(1)$ chiral symmetry, but it is consistent with the non-invertible symmetry uncovered in Refs.~~\cite{Choi:2022jqy,Cordova:2022ieu}.

\section{Useful results in 2d YM}\label{sec:2dYM}
\begin{figure}[h]
\centering
\includegraphics[width=1\textwidth]{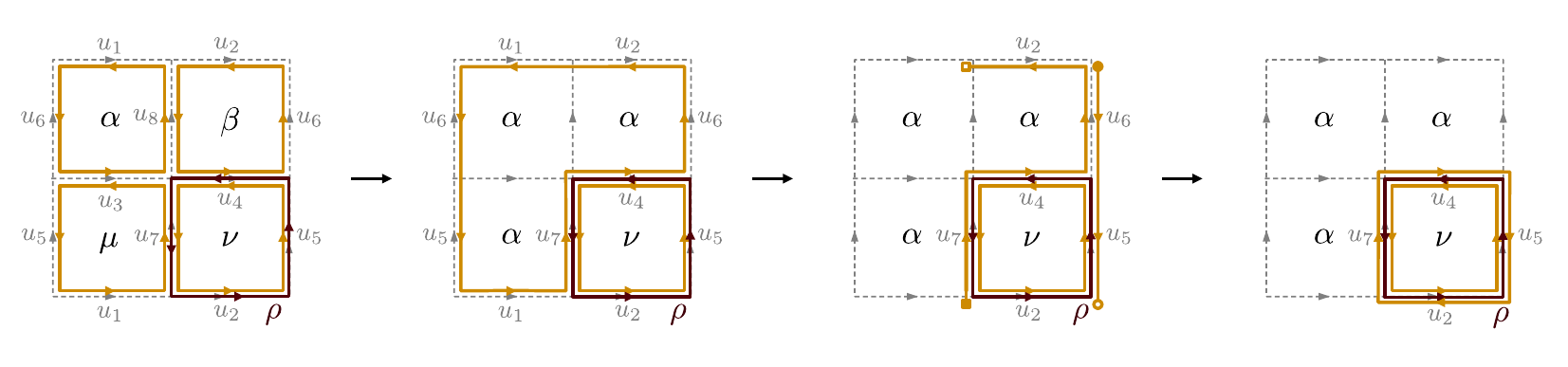}
\caption{The Wilson loop expectation value on the torus in pure 2d YM can be computed on a lattice consisting of four plaquettes. The link variable integrations required to get from one picture to the next are described in the text.}
\label{fig:4plaquette}
\end{figure}

In this appendix we provide some details on calculations in pure 2d YM theory which are necessary for the hopping expansion in 2d QCD. We use the heat-kernel formulation
\begin{equation}
Z = \prod_\ell \int du_\ell \, \prod_p \sum_\alpha d_\alpha\chi_\alpha(u_p)\, e^{-g^2 c_\alpha A_p} \,.
\end{equation}
Simple calculations in 2d YM can be performed using the following group integration formulas, 
\begin{align}
\int du\, \chi_\alpha(u_1\, u)\chi_\beta(u^\dagger u_2) &= \delta_{\alpha\beta}\frac{\chi_\alpha(u_1u_2)}{d_\alpha}\,, \\
\int du\, \chi_\alpha(u\, u_1\, u^\dagger u_2) &= \frac{\chi_\alpha(u_1)\chi_\alpha(u_2)}{d_\alpha}\,.
\end{align}
Using the fact that the heat-kernel action is subdivision invariant \cite{Migdal:1975zg,Drouffe:1978py,Lang:1981rj,Menotti:1981ry,Witten:1991we}, we can perform computations on the smallest possible lattices. For instance, the partition function on the torus (with periodic boundary conditions) can be easily computed using a single plaquette, 
\begin{align}
Z &= \int du_1 \int du_2 \, \sum_\alpha d_\alpha \chi_\alpha(u_1 u_2u_1^\dagger u_2^\dagger) e^{-g^2c_\alpha A} \\
&= \int du_2\, \sum_\alpha d_\alpha \frac{\chi_\alpha(u_2)\chi_\alpha(u_2^\dagger)}{d_\alpha}e^{-g^2c_\alpha A} = \sum_\alpha d_\alpha \frac{\chi_\alpha(\mathbbm{1})}{d_\alpha^2}e^{-g^2c_\alpha A} = \sum_\alpha e^{-g^2c_\alpha A}\,.
\end{align}
In the infinite volume limit $A \to \infty$, the only surviving term comes from $\alpha = \mathbbm{1}$ since $c_{\mathbbm{1}}=0$, and so $Z \to 1$. Similarly, Wilson loop expectation values can be evaluated on a lattice of four plaquettes as in Fig.~\ref{fig:4plaquette}. The numerator of the expectation value is
\begin{align}
\ldangle W_\rho(C) \rdangle = \prod_{\ell=1}^8 \int du_\ell \,\chi_\rho(u_2 u_5 u_4^\dagger u_7^\dagger)
 \sum_{\alpha,\beta,\mu,\nu} e^{-g^2 (c_\alpha+c_\beta+c_\mu) \frac{A-A[C]}{3}}e^{-g^2 c_\nu A[C]} \nonumber \\
\times \, d_\alpha\chi_\alpha(u_3u_8u_1^\dagger u_6^\dagger)d_\beta\chi_\beta(u_4 u_6 u_2^\dagger u_8^\dagger) d_\mu\chi_\mu(u_1 u_7 u_3^\dagger u_5^\dagger) d_\nu \chi_\nu(u_2 u_5 u_4^\dagger u_7^\dagger)\,, 
\end{align}
where we have distributed the area lying outside of the Wilson loop equally among the three plaquettes.
%%%%%%%%%%
%%%%%%%%%%%%%%

Integrating over $u_3$ and $u_8$ sets $\alpha = \beta = \mu$, so that 
\begin{align}
\ldangle W_\rho(C) \rdangle = \prod_{\ell=1,2,4,5,6,7} \int du_\ell \,\chi_\rho(u_2 u_5 u_4^\dagger u_7^\dagger)
 \sum_{\alpha,\nu} e^{-g^2 c_\alpha (A-A[C])}e^{-g^2 c_\nu A[C]} \nonumber \\
\times \, d_\alpha\chi_\alpha(u_1u_7u_4u_6u_2^\dagger u_1^\dagger u_6^\dagger u_5^\dagger) d_\nu \chi_\nu(u_2 u_5 u_4^\dagger u_7^\dagger)\,. 
\end{align}
Integrating over $u_1$ and then $u_6$ yields
\begin{align}
\ldangle W_\rho(C) \rdangle = \prod_{\ell=2,4,5,6,7} \int du_\ell \,&\chi_\rho(u_2 u_5 u_4^\dagger u_7^\dagger)
 \sum_{\alpha,\nu} e^{-g^2 c_\alpha (A-A[C])}e^{-g^2 c_\nu A[C]} \nonumber \\
\times \, &\chi_\alpha(u_7u_4u_6u_2^\dagger)\chi_\alpha(u_6^\dagger u_5^\dagger) d_\nu \chi_\nu(u_2 u_5 u_4^\dagger u_7^\dagger)\,. \\
= \prod_{\ell=2,4,5,7} \int du_\ell \,&\chi_\rho(u_2 u_5 u_4^\dagger u_7^\dagger)
 \sum_{\alpha,\nu} e^{-g^2 c_\alpha (A-A[C])}e^{-g^2 c_\nu A[C]} \nonumber \\
\times \, &\frac{1}{d_\alpha}\chi_\alpha(u_7u_4u_5^\dagger u_2^\dagger) d_\nu \chi_\nu(u_2 u_5 u_4^\dagger u_7^\dagger)\,. 
\end{align}
The steps leading to this result are depicted in the sequence in Fig.~\ref{fig:4plaquette}. The remaining integral reduces to 
\begin{align}
\int du \, \chi_\alpha(u) \chi_\nu(u^\dagger)\chi_\rho(u^\dagger) = \sum_{\gamma} \int du\, \chi_\alpha(u) N_{\nu\rho}^\gamma \chi_\gamma(u^\dagger) = N_{\nu\rho}^\alpha \,,
\end{align}
where $N_{\nu\rho}^\alpha$ is the multiplicity of $\alpha$ in $\nu\otimes\rho$. Hence, using the fact that $N_{\nu\rho}^\alpha = N_{\alpha\overline{\rho}}^\nu = N_{\overline{\alpha}\rho}^{\overline{\nu}}$ and that $d_{\alpha} = d_{\overline{\alpha}}, c_\alpha = c_{\overline{\alpha}}$, 
\begin{align} \label{eq:2dwilson}
\ldangle W_\rho(C) \rdangle = \sum_{\alpha,\nu} \frac{d_\nu}{d_\alpha} N_{\alpha\rho}^\nu \, e^{-g^2 c_\alpha (A-A[C])}e^{-g^2 c_\nu A[C]}\,.
\end{align}
In the infinite volume limit $A \to \infty$ with $A[C]$ held fixed, the only surviving term comes from $\alpha = \mathbbm{1}$, and we find
\begin{align}
\langle W_\rho(C) \rangle = d_\rho \, e^{-g^2 c_\rho A[C]}\,. 
\end{align}
There is a simple rule that allows one to immediately write down the result \eqref{eq:2dwilson} and most correlation functions of Wilson loops. Each Wilson loop insertion separates spacetime into two regions. Each region is assigned a representation $\alpha$, which is summed over, and contributes to the correlator a factor of $d_\alpha^{\chi} e^{-g^2 c_\alpha A_r}$, where the Euler characteristic $\chi = 2-2\mathfrak g - b$ is determined by the genus $\mathfrak g$ and number of boundaries $b$ of the region, which has area $A_r$. For two regions separated by a Wilson loop in representation $\rho$, one inserts the factor $N_{\alpha\rho}^\beta$, where $\alpha$, $\beta$ are the representations assigned outside and inside of the Wilson loop, respectively.  In  \eqref{eq:2dwilson}, the region inside the Wilson loop is a disk with $\mathfrak g = 0, b = 1$ and $\chi = 1$, so it contributes $d_{\nu}e^{-g^2 c_{\nu} A[C]}$, the region outside is a torus with one puncture, so it has $ \mathfrak g = 1, b=1$, and $\chi = -1$, and contributes $d_{\alpha}^{-1} e^{-g^2 c_{\alpha} (A - A[C])}$. Finally, the fact that these regions are separated by a Wilson loop in representation $\rho$ yields the factor of $N_{\alpha \rho}^{\nu}$ in \eqref{eq:2dwilson}.

%%%%%%%%%%%%%%%%%%%%%%%%%% 
\begin{figure}[h!] 
\centering
\subfigure[]{\includegraphics[width=0.3\textwidth]{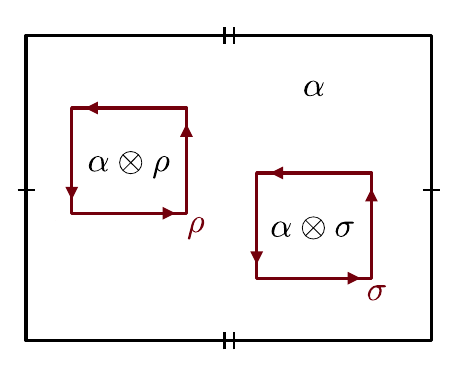}
\label{fig:two_wilson}}
\hspace{1cm}
\subfigure[]{\includegraphics[width=0.3\textwidth]{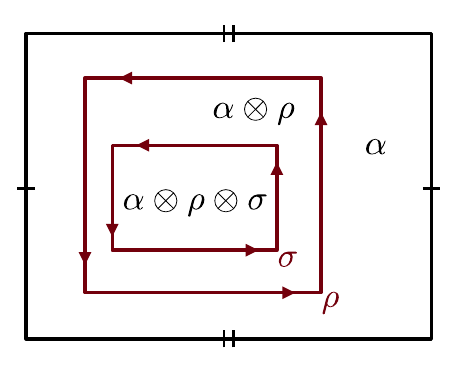}
\label{fig:concentric_wilson}}
\caption{Two simple correlation functions of Wilson loops. Each Wilson loop separates spacetime into disjoint regions, which are assigned representations as indicated.    }
\end{figure}
%%%%%%%%%%%%%%%%%%%%%%%%%%

As another example, we can use the rules above to show that for two separated non-intersecting Wilson loops on contours $C$ and $C'$ as in Fig.~\ref{fig:two_wilson}, we have
\begin{align}
\ldangle W_\rho(C) W_\sigma(C')\rdangle = \sum_{\alpha,\mu,\nu} \frac{d_\mu d_\nu}{d_\alpha^2} N_{\alpha \rho}^\mu N_{\alpha\sigma}^\nu e^{-g^2(c_\mu-c_\alpha) A[C]} e^{-g^2 (c_\nu-c_\alpha) A[C']} e^{-g^2 c_\alpha A}\,.
\end{align}
Again, in the infinite volume limit the above expression simplifies, 
\begin{equation}
\langle W_\rho(C) W_\sigma(C')\rangle = d_\rho d_\sigma \, e^{-g^2c_\rho A[C]} e^{-g^2c_\sigma A[C']} = \langle W_\rho(C)\rangle\langle W_\sigma(C')\rangle\,.
\end{equation}
For concentric loops with $C'$ inside $C$ as in Fig.~\ref{fig:concentric_wilson}, we instead find
\begin{align}
\ldangle W_\rho(C) W_\sigma(C')\rdangle = \sum_{\alpha,\mu,\nu} \frac{d_\nu}{d_\alpha} N_{\alpha \rho}^\mu N_{\mu\sigma}^\nu e^{-g^2(c_\mu-c_\alpha) A[C]} e^{-g^2 (c_\nu-c_\mu) A[C']} e^{-g^2 c_\alpha A}\,,
\end{align}
which in the infinite volume limit becomes
\begin{align}
\ldangle W_\rho(C) W_\sigma(C')\rdangle = \sum_{\nu} d_\nu  N_{\rho\sigma}^\nu e^{-g^2c_\rho A[C]} e^{-g^2 (c_\nu-c_\rho) A[C']}\,.
\end{align}
Taking $\rho =\F, \sigma = \overline{\F}$ and $C' \to C$ gives a correlation function relevant for the screening of a fundamental Wilson loop in the hopping expansion, 
\begin{align}
\langle W_{\F}(C)W_{\overline{F}}(C) \rangle = 1 + (N^2-1) e^{-g^2 c_{\text{adj}} A[C]}\,. 
\end{align}

\begin{figure}[th] 
\centering
\subfigure[]{\includegraphics[width=0.3\textwidth]{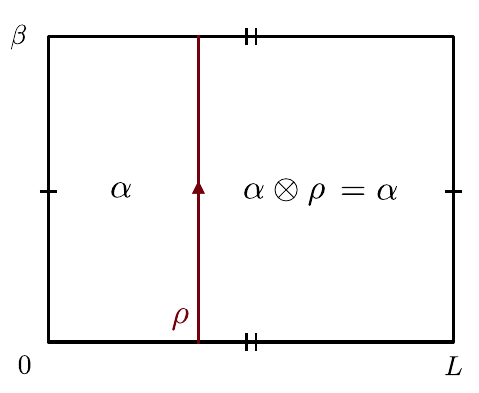}
\label{fig:one_polyakov}}
\hspace{1cm}
\subfigure[]{\includegraphics[width=0.3\textwidth]{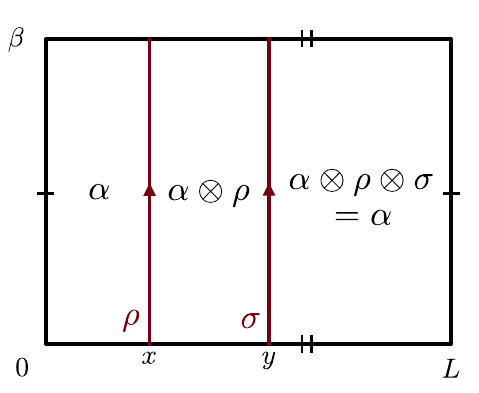}
\label{fig:two_polyakov}}
\caption{ The one- and two-point functions of Polyakov loops on the torus in pure 2d YM. Periodicity of the torus forces one to identify the representations appearing in the regions to the left and right of the Polyakov loop(s).  }
\end{figure}

Applying the above logic to the Polyakov loop, which cuts the spacetime torus open into a cylinder with two boundaries as in Fig.~\ref{fig:one_polyakov}, one finds
\begin{align}
\ldangle \tr_\rho P \rdangle = \sum_\alpha N_{\alpha\rho}^\alpha \, e^{-g^2 c_\alpha A}\,,
\end{align}
which vanishes identically if $n_\rho \not=0$. If $n_\rho = 0$, and $\rho\not=\mathbbm{1}$, the above expression is finite in finite volume. Similarly, the two-point function of Polyakov loops is
\begin{align}
\ldangle \tr_\rho P(x) \tr_\sigma P(y) \rdangle = \sum_{\alpha,\gamma} N_{\alpha\rho}^\gamma N_{\gamma\sigma}^\alpha \, e^{-g^2 (c_\gamma -c_\alpha) \beta |x-y|} e^{-g^2 c_\alpha A}\,,
\end{align}
where $\beta$ is the length of the cycle around which the Polyakov loops are wrapped, see Fig.~\ref{fig:two_polyakov}. 

%%%%%%%%%%%%%%%%%%%%%%%%%% 
%%%%%%%%%%%%%%%%%%%%%%%%%%

The expectation values of the topological operators generating the 1-form symmetry of 2d YM can be computed on a lattice consisting of a single plaquette. For simplicity we only consider the $\ZZ_N$ symmetry operators --- for a thorough discussion of the non-invertible symmetry operators see \cite{Nguyen:2021naa}. Repeating the calculation of the partition function but in the presence of the operator $U_k(\tilde x)$, 
\begin{align}
 \ldangle U_k(\tilde x)\rdangle &= \int du_1 \int du_2 \, \sum_\alpha d_\alpha \chi_\alpha(u_1 u_2u_1^\dagger u_2^\dagger \omega^{-k}) e^{-g^2c_\alpha A} \\
&= \int du_2\, \sum_\alpha d_\alpha \frac{\chi_\alpha(u_2)\chi_\alpha(u_2^\dagger\omega^{-k})}{d_\alpha}e^{-g^2c_\alpha A} \\
&= \sum_\alpha d_\alpha \frac{\chi_\alpha(\omega^{-k})}{d_\alpha^2}e^{-g^2c_\alpha A} = \sum_\alpha e^{\frac{2\pi ik}{N} n_{\alpha}} e^{-g^2c_\alpha A}\,.
\end{align} 
In the infinite volume limit, $\langle U_k(\tilde x) \rangle = 1$. More generally, the insertion of a 1-form symmetry operator in a correlation function produces an extra factor of $e^{\frac{2\pi ik}{N} n_{\alpha}}$, where $\alpha$ is the representation assigned to the region where the operator is inserted. For instance, if $\tilde x$ ($\tilde y$) lies inside (outside) a Wilson loop contour $C$, 
\begin{align}
\ldangle W_\rho(C) U_k(\tilde x) \rdangle &=  \sum_{\alpha,\nu} \frac{d_\nu}{d_\alpha} N_{\alpha\rho}^\nu e^{\frac{2\pi ik}{N} n_{\nu}} \, e^{-g^2 c_\alpha (A-A[C])}e^{-g^2 c_\nu A[C]}\,, \\
\ldangle W_\rho(C) U_k(\tilde y) \rdangle &=  \sum_{\alpha,\nu} \frac{d_\nu}{d_\alpha} N_{\alpha\rho}^\nu e^{\frac{2\pi ik}{N} n_{\alpha}} \, e^{-g^2 c_\alpha (A-A[C])}e^{-g^2 c_\nu A[C]}\,.
\end{align}
The $N$-alities are additive $n_\nu = n_\alpha + n_\rho$, and as a result $\langle W_\rho(C) U_k(\tilde x) \rangle = e^{\frac{2\pi ik}{N} n_{\rho} } \langle W_\rho(C) U_k(\tilde y) \rangle$.

\section{Center-vortex--Wilson loop correlation functions}
\label{sec:vortex_correlator_appendix}

For completeness, we compute the correlation functions of the normalized center-vortex operators $V_k$ with open and closed Wilson lines to low orders in the hopping expansion in 2d scalar QCD. First, consider the straight open Wilson line connecting the sites $0$ and $x$ on the lattice. Suppose we insert a center-vortex operator at the point $\tilde y$ on the dual lattice, which is located a transverse distance $r$ from the Wilson line, with $r$ a half-integer. The correlation function
\begin{equation}
\left \langle \frac{1}{N}\phi^\dagger(x) W_{x,0}\phi(0)\, V_k(\tilde y) \right\rangle
\end{equation} 
is $\O(1)$ in the large $N$ limit, as follows. The leading contribution to the above correlation function consists of a single, straight hopping line connecting the sites $0$ and $x$. The result is the same as in Eq.~\eqref{eq:openline}, 
\begin{equation}
\label{eq:openline_appendix}
\frac{1}{m^2}\left(\frac{\kappa}{m^2}\right)^{|x|} =\frac{1}{m^2} e^{-\mu|x|}\,.
\end{equation}
Corrections to this leading order result come from small fluctuations in the hopping line connecting the endpoints of the Wilson line insertion. A disconnected hopping loop which encircles the point $\tilde y$ does not lead to $k$-dependence in the above correlator, due to cancellation against the denominator implicit in the definition of $V_k$. Hence, the leading contribution in the hopping expansion which is sensitive to the presence of the defect occurs at $\O(\kappa^{|x|+2r+1})$. At this order, there is a hopping loop connecting the points $0$ and $x$ that makes a `detour' around the point $\tilde y$. It is easy to see that such a contribution, which is $r$-dependent and hence non-topological, is $\O(1)$ at large $N$ and therefore unsuppressed compared to the $r$- and $k$-independent terms. This corroborates the general argument based on endability. 

%%%%%%%%%%
\begin{figure}[h]
\centering
\includegraphics[width=1\textwidth]{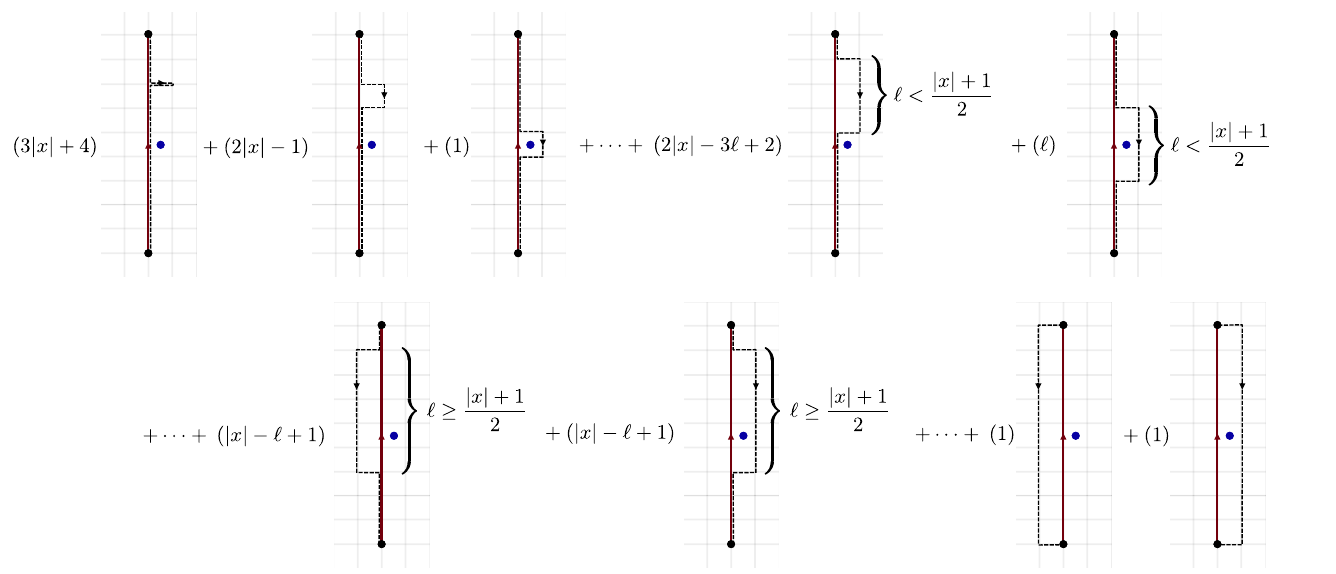}
\caption{Leading corrections     in the hopping expansion to the correlation function of an open Wilson line and center-vortex operator \eqref{eq:openlinecorrelator}. The dashed lines are the hopping lines, representing the worldlines of heavy charged matter. Here we took the length of the open Wilson line $|x|$ to be odd, and placed the center-vortex operator as close as possible to the midpoint of the line. The correlation function depends on the precise location of the center-vortex in relation to the Wilson line, even at large $N$.}
\label{fig:openlinecorrelator}
\end{figure}
%%%%%%%%%%%%%%

To see this from a concrete calculation, consider the case $r = \tfrac{1}{2}$, so that the defect is as close as possible to the open Wilson line, and suppose that the defect is placed next to the mid-point of the line. In this case the combinatorics are sufficiently simple that the calculation is straightforward.  The relevant diagrams at order $\kappa^{|x|+2}$ and their associated multiplicities are shown in Fig.~\ref{fig:openlinecorrelator}.  (Note also that to this order, the correlation function is the same regardless of whether we use the unnormalized operator $U_k$.) Summing these contributions, we find
\begin{align}\label{eq:openlinecorrelator}
\left \langle \frac{1}{N}\phi^\dagger(x) W_{x,0}\,\phi(0)\, V_k(\tilde y) \right \rangle =& \\
\frac{1}{m^2} e^{-\mu|x|}\Bigg[1 +&\left(\frac{\kappa}{m^2}\right)^2 \left(\frac{(3|x|+5)(|x|+3)}{4} + e^{\frac{\pm 2\pi i k}{N}} \frac{(|x|+1)^2}{4}\right) + \O(\kappa^4) \Bigg]\,,
\nonumber
\end{align}
where the $\pm$ depends on whether $\tilde y$ is to the right or left of the Wilson line. If we move the defect one lattice unit farther from the Wilson line and work to the same order in the hopping expansion, the result is 
\begin{align}
\left\langle \frac{1}{N}\phi^\dagger(x) W_{x,0}\,\phi(0)\, V_k(\tilde y') \right \rangle &= \frac{1}{m^2} e^{-\mu|x|}\left[ 1 + \left(\frac{\kappa}{m^2}\right)^2 (|x|+2)^2 + \O(\kappa^4) \right]\,,
\end{align}
with the difference between these two correlation functions being $\O(1)$. 

Despite the clear non-topological nature of $V_k$ in the presence of open Wilson lines at large $N$, it is still interesting to work out what happens in correlation functions with closed Wilson loops. The large-$N$ scaling argument given in Section~\ref{sec:quark_loop} implies that quark loops are unsuppressed in generic correlation functions of the unnormalized center vortex operators $U_k$. An attractive alternative is to use the normalized operators $V_k$ which have unit expectation values --- one might hope that this is sufficient to ensure topological correlation functions. However, both endability and explicit computations using the hopping expansion on the lattice indicate that even the rescaled operators do not have topological correlation functions in general at large $N$. This does not, however, rule out the possibility that \emph{some} correlation functions of $V_k$ depend only on topological data at large $N$. We now turn to an intriguing apparent example of such a correlation function, namely 
\begin{equation}
\left\langle V_k(\tilde x)\, \tfrac{1}{d_{\! R}}W_R(C) \right\rangle \quad \tilde x \text{ inside } C. 
\end{equation}

\begin{figure}[h]
\centering
\includegraphics[width=1\textwidth]{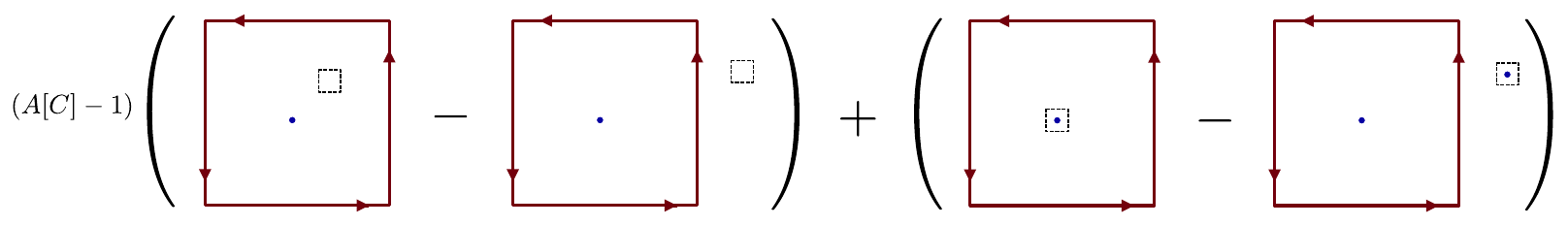}
\caption{Leading corrections     in the hopping expansion to the correlation function \eqref{eq:vortexwilson1} of a closed contractible Wilson loop and normalized center-vortex operator. The dashed lines are hopping loops, which run in both orientations around the single plaquette. }
\label{fig:vortexwilson}
\end{figure}

The first corrections to the pure gauge theory result come from summing over single plaquette hopping loops. In the numerator of the above expectation value, hopping loops can be outside the Wilson loop, inside the Wilson loop but not encircling the center-vortex, or inside the Wilson loop and encircling the center-vortex. The relevant contributions are depicted in Fig.~\ref{fig:vortexwilson}, and result in 
\begin{align} \label{eq:vortexwilson1} 
\left\langle V_k(\tilde x)\, \tfrac{1}{d_{\! R}} W_R(C) \right\rangle = e^{\frac{2\pi i k}{N}n_{\!R}} \left\langle\tfrac{1}{d_{\! R}} W_R(C)\right\rangle& \\
+\left(\frac{\kappa}{m^2}\right)^4 e^{\frac{2\pi i k}{N}n_{R}} e^{-g^2c_{R}A[C]}&\left[\sum_\rho \frac{d_\rho}{d_{\! R}} e^{-g^2(c_\rho-c_{R})}(N_{R\F}^\rho e^{\frac{2\pi i k}{N}}  + N_{R\overline{\F}}^\rho e^{-\frac{2\pi i k}{N}})\right. \nonumber \\
&\quad - N e^{-g^2c_{\F}} (e^{\frac{2\pi i k}{N}}+e^{-\frac{2\pi i k}{N}}) \nonumber \\
&\quad - \left.\sum_\rho \frac{d_\rho}{d_{\! R}} e^{-g^2(c_\rho-c_{R})}(N_{R\F}^\rho  + N_{R\overline{\F}}^\rho ) + 2N e^{-g^2c_{\F}} \right]\,. \nonumber 
\end{align}
The second, third, and fourth lines are each generically $\O(N)$. However, significant cancellation occurs so that the net result for the sum of these terms is actually $\O(1/N)$. 

%%%%%%%%%%
%%%%%%%%%%%%%%

To see the cancellation, let $R$ be a so-called `composite representation' in the sense of Ref.~\cite{GROSS1993181}. This means that $R$ appears in the tensor product of $n$ fundamental and $\bar n$ anti-fundamental representations, where $n,\bar n = \O(1)$, such that the dimension $d_{\! R} = \O(N^{n+\bar n})$ and $c_{ R} = \O(N)$. Such representations are precisely those whose expectation values have smooth (i.e. non-vanishing) large $N$ limits. The irreducible representations $\rho$ obtained by tensoring $R$ with $\F$ or $\overline\F$ are themselves composite and have dimensions $d_\rho = \O(N^{n+\bar n \pm 1})$. Moreover, for the subset with $d_\rho = \O(N^{n+\bar n +1})$ the Casimirs add, $c_\rho = c_{ R} + c_{\F} + \O(1/N)$. As a result, 
\begin{align}
\sum_\rho \frac{d_\rho}{d_{\!R}} e^{-g^2(c_\rho - c_R)}N_{R\F}^\rho &= \sum_{\substack{\rho, \\ d_\rho = \O(N^{n+\bar n+1})} }\frac{d_\rho}{d_{\!R}} e^{-g^2(c_\rho - c_R)}N_{R\F}^\rho + \O(1/N) \\
&= e^{-g^2c_{\F}} \sum_{\substack{\rho, \\ d_\rho = \O(N^{n+\bar n+1})} }\frac{d_\rho}{d_{\!R}} + \O(1/N) \\
&= e^{-g^2c_{\F}} N + \O(1/N),
\end{align}
neglecting exponential-in-$N$ corrections. Plugging this result into Eq.~\eqref{eq:vortexwilson1}, one finds
\begin{align} \label{eq:vortexwilson2} 
\left\langle V_k(\tilde x)\, \tfrac{1}{d_{\! R}} W_R(C) \right \rangle &= e^{\frac{2\pi i k}{N}n_{\!R}} \left \langle\tfrac{1}{d_{\! R}} W_R(C)\right\rangle + \O(1/N) \\
&=e^{\frac{2\pi i k}{N}n_{\!R}} \left \langle V_k(\tilde x')\, \tfrac{1}{d_{\! R}} W_R(C) \right\rangle + \O(1/N) \,,
\end{align}
where we have used the fact that to the order to which we are working, the second equality holds for $\tilde x'$ outside of the contour $C$. This shows that to the leading nontrivial order in the hopping expansion, the operators $V_k$ in large $N$ QCD act like their pure YM counterparts in correlation functions with closed Wilson loops. Namely, the correlation function only depends on the linking number of $\tilde x$ and $C$, with a phase difference determined by the representation of the Wilson loop. 

Given the wealth of evidence against the operators $V_k$ being topological discussed in the main text, one might wonder whether the above result is simply a first-order accident. However, we have verified in the case of the fundamental Wilson loop that the suppression holds to higher orders. Beyond $\O(\kappa^4)$, one can have single hopping loops of various shapes and sizes. Such contributions will give qualitatively similar results to the single-plaquette hopping loop. The more interesting contributions, which appear at $\O(\kappa^8)$, consist of two single-plaquette hopping loops which are either disconnected, or placed on top of each other. Taking into account these contributions, one finds
\begin{align}
\langle V_k(\tilde x)\,\tfrac{1}{N}W_{\F}(C)\rangle &= e^{\frac{2\pi ik}{N}} \langle V_k(\tilde x')\,\tfrac{1}{N}W_{\F}(C)\rangle \\
&+\left(\frac{\kappa}{m^2}\right)^4e^{-\frac{\lambda}{2}A[C]} (1-e^{\frac{2\pi ik}{N}})\frac{2}{N}\sinh(\lambda/2)\left(1 - \left(\frac{\kappa}{m^2}\right)^4\frac{2}{N}\sinh(\lambda/2) + \cdots\right)\,, \nonumber
\end{align}
again displaying the $1/N$-suppression from the pure gauge result.   

It would be interesting to find a proof that the $1/N$ suppression seen above persists to all orders in $\kappa$.  If it does (as seems plausible), then it would be interesting to find some interpretation or applications of this feature of the $V_k(\tilde{x})$ operators.  However, it is worth emphasizing that none of this would contradict our general arguments in the main text regarding the absence of a 1-form $\Z_N$ symmetry in large $N$ QCD. 

\linespread{1}\selectfont
\bibliographystyle{utphys}
\bibliography{non_inv}

\end{document}